\documentclass[10pt,preprint]{aastex}
\usepackage[utf8]{inputenc}

\usepackage[english]{babel}
\usepackage{natbib}
\setcitestyle{authoryear,open={(},close={)}}

\usepackage{rotating}

\usepackage[caption=false]{subfig}
\usepackage{array}
\usepackage{wrapfig}
\usepackage{multirow}
\usepackage{tabu}
\usepackage{graphicx}
\usepackage{import}
\usepackage{subfiles}
\usepackage{pdfpages}
\usepackage{pgffor}
\usepackage{comment}
\usepackage{xfrac}
\usepackage{xifthen}
\usepackage{color}
\usepackage[normalem]{ulem}
\usepackage{csquotes}
\usepackage{amsmath}
\usepackage{comment}

\newcommand{\farcsec}{\hbox{$.\!\!^{\prime\prime}$}}
\newcommand{\farcmin}{\hbox{$.\!\!^{\prime}$}}

\newcommand{\Liv}{1}
\newcommand{\UH}{2}
\newcommand{\MIT}{3}
\newcommand{\UCSC}{4}
\newcommand{\Ames}{5}
\newcommand{\Dunlap}{6}
\newcommand{\GemChile}{7}
\newcommand{\Berk}{8}
\newcommand{\UCSD}{9}
\newcommand{\Kavli}{10}
\newcommand{\NRC}{11}
\newcommand{\Vic}{12}
\newcommand{\Seti}{13}
\newcommand{\LAM}{14}

\newcommand{\bestsigmarel}{0.8}

\newcommand{\bestfitpaeso}{$319\pm6^{\circ}$} 
\newcommand{\bestfitsepeso}{$1801\pm200$} 
\newcommand{\bestredchisqeso}{$0.45$}
\newcommand{\bestdofeso}{$112$}

\newcommand{\bestmtot}{$62.0^{+2.0}_{-1.7}$}

\newcommand{\massratio}{$0.82^{+0.03}_{-0.03}$}

\newcommand{\pmra}{$-2.76216^{+0.00243}_{-0.00243}$}
\newcommand{\pmdec}{$0.35779^{+0.00344}_{-0.00344}$}
\newcommand{\para}{$0.50101^{+0.00052}_{-0.00053}$}

\newcommand{\masspri}{$34.2^{+1.3}_{-1.1}$}
\newcommand{\masssec}{$27.9^{+1.1}_{-1.0}$}

\newcommand{\bestmtoterr}{$3.25$\%~}

\newcommand{\temp}{14} 
\newcommand{\nwalkers}{$300$} 
\newcommand{\nsteps}{$2\times10^6$}

\begin{document}

\author{
E. Victor Garcia\altaffilmark{\Liv},
S. Mark Ammons\altaffilmark{\Liv},
Maissa Salama\altaffilmark{\UH},
Ian Crossfield\altaffilmark{\MIT,\UCSC},
Eduardo Bendek\altaffilmark{\Ames},
Jeffrey Chilcote\altaffilmark{\Dunlap},
Vincent Garrel\altaffilmark{\GemChile},
James R. Graham\altaffilmark{\Berk},
Paul Kalas\altaffilmark{\Berk},
Quinn Konopacky\altaffilmark{\UCSD},
Jessica R. Lu\altaffilmark{\UH},
Bruce Macintosh\altaffilmark{\Kavli},
Eduardo Marin\altaffilmark{\GemChile},
Christian Marois\altaffilmark{\NRC,\Vic}
Eric Nielsen\altaffilmark{\Kavli,\Seti},
Benoît Neichel\altaffilmark{\LAM},
Don Pham\altaffilmark{\Liv},
Robert J. De Rosa\altaffilmark{\Berk},
Dominic M. Ryan\altaffilmark{\Berk},
Maxwell Service\altaffilmark{\UH},
Gaetano Sivo\altaffilmark{\GemChile}
}

\altaffiltext{\Liv}{Lawrence Livermore National Laboratory, L-210, 7000 East Avenue, Livermore, CA 94550, USA; eugenio.v.garcia@gmail.com}

\altaffiltext{\UH}{Institute for Astronomy, University of Hawai‘i at Manoa, Honolulu, HI 96822, USA}
\altaffiltext{\MIT}{Department of Physics, Massachusetts Institute of Technology, 77 Massachusetts Avenue, Cambridge, MA 02139, USA}
\altaffiltext{\UCSC}{University of California, Santa Cruz, 1156 High St., Santa Cruz, CA 95064, USA}
\altaffiltext{\Ames}{NASA Ames Research Center, Moffett Field, CA 94035, USA}
\altaffiltext{\Dunlap}{Dunlap Institute for Astronomy and Astrophysics, University of Toronto, Toronto, ON M5S 3H4, Canada}
\altaffiltext{\GemChile}{Gemini Telescope, Colina el Pino S/N, La Serena, Chile}
\altaffiltext{\Berk}{Astronomy Department, University of California, Berkeley, CA 94720, USA}
\altaffiltext{\UCSD}{Center for Astrophysics and Space Sciences, University of California, San Diego, La Jolla, CA 92093, USA}
\altaffiltext{\Kavli}{Kavli Institute for Particle Astrophysics and Cosmology, Stanford University, Stanford, CA 94305, USA}

\altaffiltext{\NRC}{National Research Council of Canada Herzberg, 5071 West Saanich Road, Victoria, BC V9E 2E7, Canada}
\altaffiltext{\Vic}{Department of Physics and Astronomy, University of Victoria, 3800 Finnerty Road, Victoria, BC V8P 5C2, Canada}

\altaffiltext{\Seti}{SETI Institute, Carl Sagan Center, 189 Bernardo Avenue, Mountain View, CA 94043, USA}
\altaffiltext{\LAM}{Aix Marseille Universit, CNRS, LAM (Laboratoire d’Astrophysique de Marseille) UMR 7326, F-13388 Marseille, France}

\title{Individual, Model-Independent Masses of the Closest Known Brown Dwarf Binary to the Sun}

\begin{abstract}

At a distance of $\sim2$~pc, our nearest brown dwarf neighbor, Luhman 16 AB, has been extensively studied since its discovery 3 years ago, yet its most fundamental parameter -- the masses of the individual dwarfs -- has not been constrained with precision. In this work we present the full astrometric orbit and barycentric motion of Luhman 16 AB and the first precision measurements of the individual component masses. We draw upon archival observations spanning 31 years from the European Southern Observatory (ESO) Schmidt Telescope, the Deep Near-Infrared Survey of the Southern Sky (DENIS), public FORS2 data on the Very Large Telescope (VLT), and new astrometry from the Gemini South Multiconjugate Adaptive Optics System (GeMS). Finally, we include three radial velocity measurements of the two components from VLT/CRIRES, spanning one year. With this new data sampling a full period of the orbit, we use a Markov Chain Monte Carlo algorithm to fit a 16-parameter model incorporating mutual orbit and barycentric motion parameters and constrain the individual masses to be~\masssec~$M_{J}$ for the T dwarf and~\masspri~$M_{J}$ for the L dwarf.  Our measurements of Luhman 16 AB's mass ratio and barycentric motion parameters are consistent with previous estimates in the literature utilizing recent astrometry only.  The GeMS-derived measurements of the Luhman 16 AB separation in 2014-2015 agree closely with Hubble Space Telescope (HST) measurements made during the same epoch \citep{Bedin17}, and the derived mutual orbit agrees with those measurements to within the HST uncertainties of $0.3 - 0.4$ milliarcseconds. 

\end{abstract}


\section{Introduction}

Individual mass measurements of brown dwarfs are critical to calibrate the mass-luminosity relationship for sub-stellar objects. To date, there is only a single field brown dwarf binary with direct individual mass measurements of its components \citep[SDSS $105213.51+442255.7$AB,][]{Dupuy15}. In the sub-stellar mass regime of $\approx0.01-0.08$~$M_{\odot}$ \citep{Dieterich14,Auddy16}, there are only four other systems with model-independent, directly measured masses with uncertainties $<10\%$: The young, eclipsing brown dwarf M6+M6 binary with masses of $0.054\pm0.005$~$M_{\odot}$ and $0.034\pm 0.003$~$M_{\odot}$ 2MASS J05352184-0546085AB from \cite{Stassun06}; a sub-stellar ($0.063\pm0.005$~$M_{\odot}$) companion GL 802B with a directly observed mass via barycentric astrometry from \cite{Ireland08}; a sub-stellar companion to HR 7672B with directly observed masses of $0.0655\pm0.0029$~$M_{\odot}$ and $0.069^{+0.003}_{-0.012}$~$M_{\odot}$ from combined orbital astrometry and radial velocity measurements from \cite{Crepp12}; and GJ 569 Bab with directly observed masses of $0.073\pm0.008$~$M_{\odot}$ and $0.053\pm0.006$~$M_{\odot}$ with combined radial velocity and orbital astrometry from \cite{Konopacky10}. There are also a number of brown dwarfs with mass determinations that depend on a stellar model-derived mass of the primary (host) star \citep{Deleuil08,Anderson11,Bouchy11a,Bouchy11b,Johnson11,Siverd12,Diaz13,Montet15,Montet16}. Given that there are only seven completely model-independent mass determinations of individual brown dwarfs (as compared to more than 30 planets and about 100 or more stars) -- more measurements are sorely needed.  



At a distance of $\sim2$ pc, Luhman 16 AB (WISE J$104915.57-531906.1$) is the closest known binary brown dwarf system. Given its proximity, Luhman 16 represents a unique opportunity to directly test brown dwarf evolution models. There has been a flurry of literature on Luhman 16 AB since it was first identified as a high proper motion L/T brown dwarf binary by \cite{Luhman13}.  The binary is composed of a L$7.5$ primary and a T$0.5$ secondary \citep{Burgasser13,Kniazev13}. Both components show Li I (6707 \AA) absorption, confirming Luhman 16 AB as the closest brown dwarf binary discovered \citep{Faherty14}, and exhibit significant photometric and spectroscopic variability suggesting patchy clouds \citep{Gillon13,Biller13,Burgasser14,Street15,Buenzli15B,Mancini15}. \cite{Lodieu15} obtained VLT X-shooter spectroscopy of Luhman 16 AB, finding evidence for lithium, placing an upper limit on the masses of $\lesssim0.06$~$M_{\odot}$. \cite{Crossfield14} and \cite{Karalidi16} find evidence for spots, using Doppler imaging and light-curve inversion techniques, respectively. Using FORS2 data published in part by \citet{Boffin14}, \cite{Sahlmann15} derived a distance of $1.998\pm0.0004$~pc, and further confirmed the high proper motion of $\approx2\farcsec8$/year of \cite{Luhman13}.  \cite{Sahlmann15} used the FORS2/VLT $I$-band imaging to obtain barycentric motion, finding a mass ratio of $q\equiv M_B/M_A=0.78\pm0.10$ for the system.  \cite{Bedin17} use astrometry recently obtained from the Hubble Space Telescope Wide Field Camera 3 (WFC3) to constrain the system's barycentric motion and mass ratio, obtaining $q\equiv M_B/M_A=0.848\pm0.023$.  \cite{Bedin17} also obtain constraints on the orbital parameters and total mass of Luhman 16 AB, although these are relatively weak due to the limited coverage of the system's full period. 

Despite the characterization of Luhman 16 AB's surfaces, cloud structures, temperature, atmospheric composition, photometric variability, parallax, proper-motion and mass ratio over the last 3 years, we have yet to directly measure Luhman 16 AB's most fundamental parameter, the individual component masses, with precision. Previous authors have noted a degeneracy between barycentric proper motion and the mass ratio of the binary \citep{Sahlmann15}, frustrating mass measurements that rely on modern epoch data alone. In this work, we constrain the total mass $M_{\rm tot}$ by combining high precision astrometry with archival imaging data from 1984 and 1999 to sample the full orbital period.  We measure positions of the individual components of Luhman 16 at the 1984 epoch, place these data on the ICRS system using publicly available GAIA DR1 data of field reference stars \citep{gaia16a, gaia16b}, and include the 1984 positions as explicit constraints in the model fit.  We then fit a global, 16-parameter model that includes seven parameters describing the mutual orbit of the binary (total mass $M_{\rm tot} = M_A+M_B$, semi-major axis log~$a$, eccentricity $e$, longitude of ascending node~$\Omega$, argument of periastron $\omega$, time of periastron passage $\tau$ and inclination cos $i$), six parameters describing the space motion of the system barycenter (parallax $\varpi_{\rm rel}$, proper motion $\mu_{\alpha,\delta}$, position $\alpha_0$, $\delta_0$, and mass ratio $q$), and three parameters modeling Chromatic Differential Atmospheric Refraction effects ($\rho_{\rm fors2}$, $d_{\rm fors2}$, $\rho_{\rm gems}$).  By jointly fitting for the mutual orbit and barycentric motion parameters, we account for any correlated uncertainty between these.  We present comparisons between our measurements and those of \cite{Bedin17} in the conclusion.


\section{Data}
\label{sect:data}

The astrometry and relative radial velocity data used in this work span 31 years, roughly the orbital period previously estimated for this binary \citep{Luhman13, Mamajek13}. We obtained astrometry of Luhman 16 AB using the following facilities: GeMS \citep{Rigaut14,Neichel14} in 2014-2015; archival data from VLT/FORS2 data from 2013-2015; a photographic plate taken with the 1-meter Schmidt telescope at ESO's La Silla Observatory in 1984; and an image from the Deep Near-Infrared Survey of the Southern Sky \citep[DENIS,][]{Epch1999} in 1999. The elongated 1984 ESO $R$-band image of Luhman 16 AB, the unresolved $I$-band DENIS 1999 image, a sample reduced GeMS 2014 and a VLT/FORS2 2013 epoch are shown in Figure~\ref{fig:sampdata}. We also include archival VLT/CRIRES radial velocity measurements from 2013-2014, described at the end of this section.

\subsection{GeMS MCAO Imaging}

Observations of Luhman 16 AB and NGC 1851 were obtained with the Gemini South
telescope using the Gemini South Adaptive Optics Imager \citep[GSAOI,][]{mcgregor04,carrasco12}
and the Gemini Multi-Conjugate Adaptive Optics System \citep[GeMS,][]{Rigaut14}. Together, the two instruments deliver near-diffraction-limited images from $0.9-2.4$~$\mu$m, 
with a FoV of 85$\times$85$\farcsec$  This AO system uses five sodium Laser Guide Stars (LGSs) to correct for atmospheric turbulence and up to three Natural
Guide Stars (NGSs) brighter than $R = 15.5$ mag to compensate for
tip–tilt and plate mode variations\footnote{For details, see ``Tip-Tilt and Plate Scales modes" here: http://www.gemini.edu/sciops/instruments/gems/introduction-gems/loops-and-offloads} over a 2$^{\prime}$ field of view (FoV)
of the AO bench unit \citep[CANOPUS;][]{Rigaut14}.  With a resolution typically exceeding $0\farcsec1$ in $H$ and $K_s$, GeMS is capable of resolving Luhman 16 AB throughout its entire orbit.  

We measure the GeMS static distortion map with observations of NGC 1851 obtained as part of the GeMS Science Verification program.  Stellar positions from GeMS are compared with known stellar positions measured with the Hubble Space Telescope Advanced Camera for Surveys \citep[ACS,][]{ford98}. The distortion measurement uses two epochs of undithered frames of NGC 1851 taken one month apart, on 2012 Dec 30 and 2013 Jan 28.  Images are flat-fielded with combined twilight flats, and masked for bad pixels using IRAF's \texttt{GAREDUCE} reduction routines.  Images are not mosaicted into 4196$\times$4196 frames.  

Images of Luhman 16 AB were obtained with GeMS on the nights of 2014 February 12, April 12, April 13, May 27, December 3 and 2015 March 5 in the $J$-continuum narrow-band filter\footnote{The $J$-continuum central wavelength is $1.207\mu$m. See here: https://www.gemini.edu//sciops/instruments/gsaoi/instrument-description/filters}.  A sample image is shown in the lower-left panel of Figure~\ref{fig:sampdata}. Exposure times were 15 seconds and a position angle of 200$^{\circ}$ was used.  AO-corrected Strehl ratios as measured on Luhman 16 AB were $0.5-5$\%.  The data were reduced following the standard procedures for near–infrared imaging provided by the Gemini/GSAOI package inside IRAF \citep{Tody1986}.   Images were flat-fielded with combined twilight flats and masked for bad pixels using IRAF's \texttt{GAREDUCE} reduction routines.

\subsection{FORS2 Imaging}

We analysed FORS2 \citep{App98} images of Luhman 16 AB
taken between 2013 April 4, and 2015 February 3, that were retrieved from the
ESO archive \citep[][Programmes 291.C-5004 and 593.C-0314; PI: Boffin]{Boffin14}.  A sample image is shown in the lower-right panel of Figure~\ref{fig:sampdata}. The images were obtained in the $I$-band Bessel filter with exposures of
15 seconds.  The images were dithered in a 3$\times$3 box pattern with
1\farcsec0 steps.  We analyzed the dark images, stacked the twilight flat
field image, and flat-fielded the science images with the ESO Reflex
Data Reduction Pipeline \citep{freudling13}.  Image FWHMs varied between 
0\farcsec51 and 0\farcsec85 and the observations span 661 days. Each of 27 epochs consists of $16-42$ frames taken over 1800 seconds on average,
resulting in 825 exposures. We exclude 107 frames with PSF FWHM$>0\farcsec75$, resulting in 718 frames.  Although more archival FORS2 images are available after 2015 February 3, we choose not to analyze these because the PSFs of the individual components of Luhman 16 AB begin to overlap after this point in time in the seeing-limited FORS2 images, making precision astrometry difficult to recover.  

\subsection{ESO $R$-band Photometric Plate Image}

We make use of a scanned plate image of Luhman 16 AB taken by the ESO 1-meter Schmidt Telescope on March 5, 1984, first identified by \citet{Mamajek13}. From Figure~\ref{fig:sampdata} (upper-left), Luhman 16 AB appears well-resolved in 1984, with the brighter component in the south-east. SUPERCOSMOS \citep{hambly01} obtained a scan of ESO red plate $R$ 5562 with a plate scale of 0\farcsec675 pixel$^{-1}$.  The filter used for the photographic plate was the RG630 filter and the plate used IIIaF emulsion.  The exposure time was 120 seconds. 

\subsection{DENIS $I$-band Image}
Luhman 16 AB was identified by \cite{Luhman13} in an $I$-band ($0.83$~$\mu$m) image from DENIS on February 11th, 1999. From Figure~\ref{fig:sampdata} (upper-right), Luhman 16 AB appears unresolved in very good seeing -- implying the binary was at small projected separation at that time. The plate scale was 1\farcsec0 pixel$^{-1}$ for the $I$-band filter and $t_{\rm exp}=9$ seconds. 

\subsection{VLT/CRIRES Relative Radial Velocities \label{sec:crires}} 

The Luhman 16 AB system was observed several times by the high-resolution infrared spectrograph CRIRES \citep{Kaeufl2004} fed by Multi-Applications Curvature Adaptive optics, as part of programs 291.C-5006 and 093.C-0335 (PI Crossfield).  The former program observed Luhman 16 for five hours on UT May 05, 2013, and the latter observed for several hours each on UT May 02 and 20, 2014.  The spectra span wavelengths from 2.288--2.345 $\mu$m; the VLT adaptive optics system spatially resolved the two brown dwarfs, so the CRIRES slit was aligned along the binary position angle to observe both brown dwarfs simultaneously.  

We use standard ESO tools to reduce these data and extracted 1D spectra, following exactly the same approach laid out by \cite{Crossfield14}. Again following that work, we calibrate the extracted spectra using a forward-modeling approach \citep{Blake2007,Bean2010}.  


In chronological order, we measure a relative velocity ($v_A - v_B$) for this system of 2740, 1940, and 1850~m~s$^{-1}$, as plotted in Fig.~\ref{fig:relrv} and shown in Table~\ref{table:relrv}. Though the relative radial velocity uncertainties are quite small (tens of m~s$^{-1}$), the uncertainties are in fact dominated by systematics resulting from the inhomogeneous brown dwarf surface (and the induced nonuniformities in the brown dwarf line profiles).  We conservatively estimate the uncertainty at each of the three epochs to be 200~m~s$^{-1}$, consistent with the internal scatter in these measurements reported by \cite{Crossfield14}.

\section{Astrometry and Uncertainty Analysis\label{sect:relastro}}

The processed, Luhman 16 astrometry from GeMS, VLT/FORS2, and the ESO photographic plate are shown in Table~\ref{table:absastro}. This astrometry still contains chromatic differential atmospheric refraction (CDAR), which is dealt with as detailed in \S\ref{sec:cdarmod}. The CDAR-subtracted separations are shown in Table~\ref{table:relastro}. 

All astrometry in this work is ultimately tied to the ICRS system \textbf{ via GAIA DR1 measurements of background stars \citep{gaia16a, gaia16b}}.  The linking of reference frames is described in Section \ref{sect:trans}.  The first linking of reference frames is between a distortion-corrected FORS2 reference frame and GAIA.  The other datasets, including the ESO photographic plate and the distortion-corrected GeMS data, are linked to the FORS2 reference frame via matched background stars.  In each data set, individual frames are never stacked before measuring stellar positions; positions are only averaged over frames following correction for static distortion.  Within a particular data set, reference frames are defined by averaging positions of background stars over the entire dataset following a rough alignment, as described below in Section \ref{sect:trans}.  These unique reference frames are used when transforming from camera to camera.  Throughout this work, tangent plane coordinates are converted to spherical coordinates using the reference point 10:49:14.765 -53:18:15.998, \textbf{which is on the ICRS system}.


\subsection{GeMS Distortion Map Measurement with NGC 1851 Observations\label{sect:gems_distortion}}
In order to obtain Luhman 16 AB astrometry from GeMS data, we first obtain the distortion map over the GeMS field of view. We use an image of globular cluster NGC 1851 to derive this distortion map. We run \texttt{Starfinder} \citep{diolaiti00} on individual frames, using 20 bright, unsaturated stars distributed throughout the field to create a master PSF for each frame.  A first extraction is performed down to a 20$\sigma$ limit, secondary stars in the PSF images are subtracted, and a new master PSF free from secondary sources is created.  Stars are then re-extracted down to 7$\sigma$ with a PSF correlation threshold of $0.7$.  The \texttt{Starfinder} function \texttt{match\_coord.pro} is used to approximately register each frame to the first in each set. Frames with fewer than 1500 stars matching the reference frame's list to within 3 pixels are rejected. This step trims set 1 from 18 to 13 frames and Set 2 from 26 to 24 frames.  

To establish a set of common stars in all frames, stars falling more than 3 pixels from the reference frame's position in any of the frames are rejected from the list, leaving about 1000 stars in each set in the full GSAOI field.  Stars located within 10 pixels of any other common stars are also eliminated.  Stars within 2 pixels of a bad pixel or within 10 pixels of the edge of the 85$^{\prime\prime}\times$85$^{\prime\prime}$ field of view are rejected. If \texttt{Starfinder} splits a single star into multiple detections in single frame, those measurements are excluded, but other frames with a single detection are retained for that star.  Following these exclusions, $\approx$1200 stars remain for each set.

For each frame, a 4 parameter linear transformation is performed to transform each GeMS chip to the appropriate area on the ACS frame to enable matching between GeMS stars and ACS stars.  The deviations between the ACS and original untransformed GeMS positions are measured and recorded for each star in each frame (a total of $2\times10^5$ stellar measurements).  A global 120 parameter fourth-order static polynomial distortion pattern (30 parameters per chip) in addition to a 6 parameter linear transformation for each frame is fitted to all stellar deviation measurements using the IDL \texttt{MPFITFUN} minimization routine.  The nonlinear components of the model remain constant to within 2 mas across the full field when the position angle of the instrument is rotated and the dither position is changed by up to $30^{\prime\prime}$.

The resulting nonlinear components of the model distortion map are shown in Figure \ref{fig:GSAOI_distortion}.  These vectors have been obtained by applying the distortion correction model to a regularly spaced grid of points and fitting and subtracting a 6-parameter linear transformation to each chip.  

\subsection{GeMS Luhman 16 AB Position Measurements \label{sec:relgems}}

The GSAOI images of Luhman 16 AB were analyzed by first running \texttt{Starfinder} on individual frames.  A 16$\times$16 pixel (0\farcsec32$\times$0\farcsec32) image of Luhman A is used as the reference PSF for cross-correlation with other stars in the field.  The correlation between the Luhman 16 A and Luhman B PSFs is consistently above 0.99 and the median value over all frames is 0.994.  We establish a set of common stars in all frames using a procedure similar to that described above for the NGC 1851 analysis.  If \texttt{Starfinder} splits a single star into multiple detections in single frame, those measurements are excluded (but other frames with a single detection are retained for that star). 

The frames are distortion corrected using the distortion map measured from NGC 1851 images, as described in the section above.  Following distortion correction, the stellar positions for all epochs are transformed to the same astrometric frame with 6 parameter linear transformations measured from $\approx12$ background stars.

\subsection{FORS2 Luhman 16 AB Position Measurements \label{sec:fors2}}

We use \texttt{Starfinder} to measure stellar positions in the FORS2 chip 1 images.  We use 20 bright, unsaturated stars distributed throughout the field to create a master PSF for each frame.  A first extraction is performed down to a $20\sigma$ limit, secondary stars in the PSF images are subtracted, and a new master PSF free from secondary sources is created.  Stars are then re-extracted down to 7$\sigma$ with a PSF correlation threshold of $0.7$.  The \texttt{Starfinder} function \texttt{match\_coord.pro} is used to approximately register each frame to the first in each set.   Frames with less than 300 stars matching the reference frame's list to within 3 pixels are rejected.  

To establish a set of common stars in all frames, stars falling more than 3 pixels from the reference frame's position in any of the frames are rejected from the list, leaving about 200 stars in each set in the full FORS2 field.  Stars located within 10 pixels of any other common stars are also eliminated.  Stars within 2 pixels of a bad pixel or within 10 pixels of the edge of the field of view are rejected.  If \texttt{Starfinder} splits a single star into multiple detections in single frame, those measurements are excluded (but other frames with a single detection are retained for that star).

\subsubsection{Generation of FORS2 Static Distortion Map\label{sec:fors2_correction}}

For each dithered frame, a 6 parameter linear transformation is performed to match the FORS2 chip to the GAIA \textbf{DR1} positions using $\approx200$ background stars.  This transformation removes the linear effects of Achromatic Differential Atmospheric Refraction \citep[ADAR chp. 4,][]{Lu08}, but nonlinear chip distortions have not yet been removed.  The deviations between the GAIA positions and original untransformed FORS2 positions are measured and recorded for each star in each frame (a total of $1.3\times10^5$ stellar measurements).  A global 20 parameter third-order static polynomial distortion pattern in addition to a 6 parameter linear transformation for each frame is fitted to all stellar deviation measurements.  The 6 additional parameters for each frame fit the 9x9 dithering pattern used in the original observations.  The output of this procedure is a set of polynomial coefficients that can be used to correct the nonlinear components of FORS2 camera distortion.



\subsection{ESO $R$-band Plate Position Measurements \label{sec:eso1984_positions}}

We run \texttt{Starfinder} on the digitized ESO 1984 $R$-band plate to detect background stars.   A 5$\times$5 pixel (3\farcsec4$\times$3\farcsec4) image of a bright, unsaturated star is used as the reference PSF for cross-correlation with other stars in the field.  As expected for a digitized photographic plate with partially saturated background stars, correlations between stars and the reference PSF are not high.  The median correlation is $0.85$ and the threshold for inclusion is $0.6$.  

\subsection{Transforming GeMS, FORS2, and ESO $R$-band Plate Astrometry to a Common Reference Frame \label{sect:trans}}

Reference frames for each camera (GeMS and FORS2) are established according to the following procedure.  First, following distortion correction on the position measurements, each frame is roughly aligned to the highest SNR frame with two shift parameters and a rotation angle. Positions of reference stars are then averaged to define that camera's unique reference frame.  For the ESO $R$-band plate, there is only one image in the data set, so the raw positions of the background stars define that reference frame.

The GeMS and ESO $R$-band data sets are then transformed to the FORS2 reference frame and then GeMS, FORS2, and ESO $R$-band are transformed to the GAIA \textbf{DR1} ICRS system \citep{gaia16a, gaia16b}.  The GeMS system is transformed to the FORS2 system with a 6-parameter linear transformation of the general form of Equation~\ref{eqn:linxform}) via matching of $14$ isolated background stars ($15 < R < 18$).  The per-coordinate standard deviation of the residuals of that fit is $2.6$ mas.  These low residuals suggest that the high order camera distortion corrections for both the GeMS and FORS2 instruments are accurate to within $2-3$ mas over the $1\farcmin4$ field of view of the GeMS camera. 

Next, we remove ADAR from our FORS2 reference frame. Rather than calculate a theoretical ADAR as a function of zenith angle, temperature, pressure, relative humidity, and wavelength of observation \citep{Gubler98,Lu08}, we transform the FORS2 system to the ADAR-free GAIA \textbf{DR1} ICRS system \citep{gaia16a, gaia16b}, using a 6-parameter linear fit:
\begin{equation}\label{eqn:linxform}
\begin{split}
x^{\prime} = & Ax + By + C \\
y^{\prime} = & Dx + Ey + F
\end{split}
\end{equation}
where the coefficients $A,B,C,D$ account for the linear ADAR and a measurement of the FORS2 plate scale, and parameters $E$ and $F$ account for the absolute shift and rotation in coordinates between the FORS2 system and the GAIA \textbf{DR1} ICRS system. Thus, we empirically remove the linear ADAR from our FORS2 and GeMS observations of Luhman 16 AB.  The quadratic component of ADAR is negligible in our case. We transformed the FORS2 system to the GAIA \textbf{DR1} ICRS with a 6-parameter linear transformation above fitted to 194 matching isolated stars ($15 < R < 18$). The standard deviation of the residuals of that fit is $1.2$ mas. Following the transformations derived from background star positions, the Luhman 16 AB separations as measured by FORS2 and GeMS agree to within the FORS2 error bars ($1-2$ mas) for comparable epochs. 

The ESO R plate is linked to the FORS2 system via a transformation consisting of a pair of 2nd-order polynomials.  These polynomials are fitted to a set of 27 common background stars within a $1\farcmin0$ radius of Luhman 16 AB.  This single step accounts for DAR and camera distortion in the ESO R plate.  The standard deviation of the residuals about that fit is 84 mas. 


\subsection{ESO $R$-band Plate Orbital Astrometry \label{sec:eso1984}}

Given the long time baseline between the 1984 ESO Schmidt photometric plate image of Luhman 16 AB and our 2013-2015 VLT/FORS2 and GeMS data, the position of either star in 1984 plays a critical role in breaking the mass-period-eccentricity degeneracies in our orbit model. Therefore, we paid special attention to deriving a separation and position angle for Luhman 16 AB from our ESO 1984 plate. 

First, we determined that the best PSF model using background stars to be a Lorentzian PSF. We fit Moffat, Gaussian and Lorentzian PSF functions to isolated, unsaturated stars in the plate image $\lesssim2^{\prime}$ of Luhman 16 AB, finding Lorentzian PSFs to yield the lowest ``model-data" residuals for fits to each nearby background star. The PSF models are constructed at $10\times$ the $0\farcs67$ pixel$^{-1}$ plate scale.  
 
Next, to model the binary Luhman 16 AB, we fit a binary model of two co-added, shifted, scaled Lorenztian PSFs to the 1984 ESO image of Luhman 16 AB  using a variant of IDL's \texttt{MPFITFUN}, which minimizes the $\chi^{2}$ using a Levenberg-Marquardt implementation in IDL \citep[\texttt{MPFIT},][]{markwardt09}. We fit for 9 free parameters: the sky background, the fluxes of Luhman 16 A and B, the positions $X,Y$ of Luhman 16 A and B on the photometric plate, and the X axis and Y axis Lorentzian widths of the Lorentzian PSFs.  We fit our binary model to a $11\times11$ pixel ($7\farcsec37\times7\farcsec37$) region centered on Luhman 16 AB. We masked a few pixels on the upper right hand corner of the image that contained a nearby star (Figure~\ref{fig:1984resid}). 

The above procedure produces a $\chi^{2}$ map of parameter space for each binary model as function of parameters of interest. Our final contour map of $\Delta\chi^{2}=\chi^{2}-\chi^{2}_{\rm min}$, is shown in Figure~\ref{fig:conts} (left panel) as dark green, medium green and light green solid lines, for $1\sigma$, $2\sigma$, and $3\sigma$ contours corresponding to $\Delta\chi^{2} < 10.4$, $17.2$, and $25.2$ respectively \citep{Press02}. 

We compare our minimum-variance binary Lorentzian Model PSF model to our 1984 ESO image in Figure~\ref{fig:1984resid}. We find the median `data-model' residuals are~$7.5\%$ (bottom-right panel, Figure~\ref{fig:1984resid}). Our minimum-variance model gives a reduced chi-square of $\chi^{2}_{\rm red}=$\bestredchisqeso~for~\bestdofeso~degrees of freedom.  Our minimum-variance PA~\bestfitpaeso~matches within uncertainty the $138^{\circ}$ elongation reported by \cite{Mamajek13}, albeit with a $180^{\circ}$ degeneracy. Our minimum-variance separation is~\bestfitsepeso~mas.

We took several steps to ensure our minimum-variance binary PSF model was not a local optimum in $\chi^{2}$ space. We ran \texttt{MPFITFUN} twice with a maximum of 3000 iterations, with the second run starting at the minimum-variance parameters of the first run. We tested the sensitivity of our minimum-variance binary PSF model to the initial guess performing $10^6$ \texttt{MPFITFUN} runs as detailed above, with random initializations for each initial guess. \texttt{MPFITFUN} converged to our best fit parameters the majority of the time, meaning our minimum-variance solution is robust to the initial guess.

The south-eastern component of the binary was identified as Luhman 16 A with a spectral type of L$7.5$ \citep{Luhman13,Kniazev13,Burgasser14}. We identify Luhman 16 B as the north-western component of the binary, which has a spectral type of T$0.5$ \citep{Luhman13,Kniazev13,Burgasser14}. Similarly, the brighter south-eastern component of the binary in 2013-2014 identified by \cite{Street15} in a $700-950$~nm filter (cf. Figure 2). We find the south-east component of the binary in the ESO Schmidt plate to have a flux ratio $F_1/F_2\approx1.2$ relative to the north-west component of the binary at $R$-band. Thus we conclude that the orientation of the binary in the 1984 epoch is the same as in 2013, i.e., Luhman 16 A is the south-eastern position and Luhman 16 B is the north-western position.

\subsection{$I$-band DENIS 1999 Orbital Astrometry Upper Limit\label{sec:denis99}}

\begin{figure}[ht!]
\begin{center}
	\includegraphics[width = \textwidth]{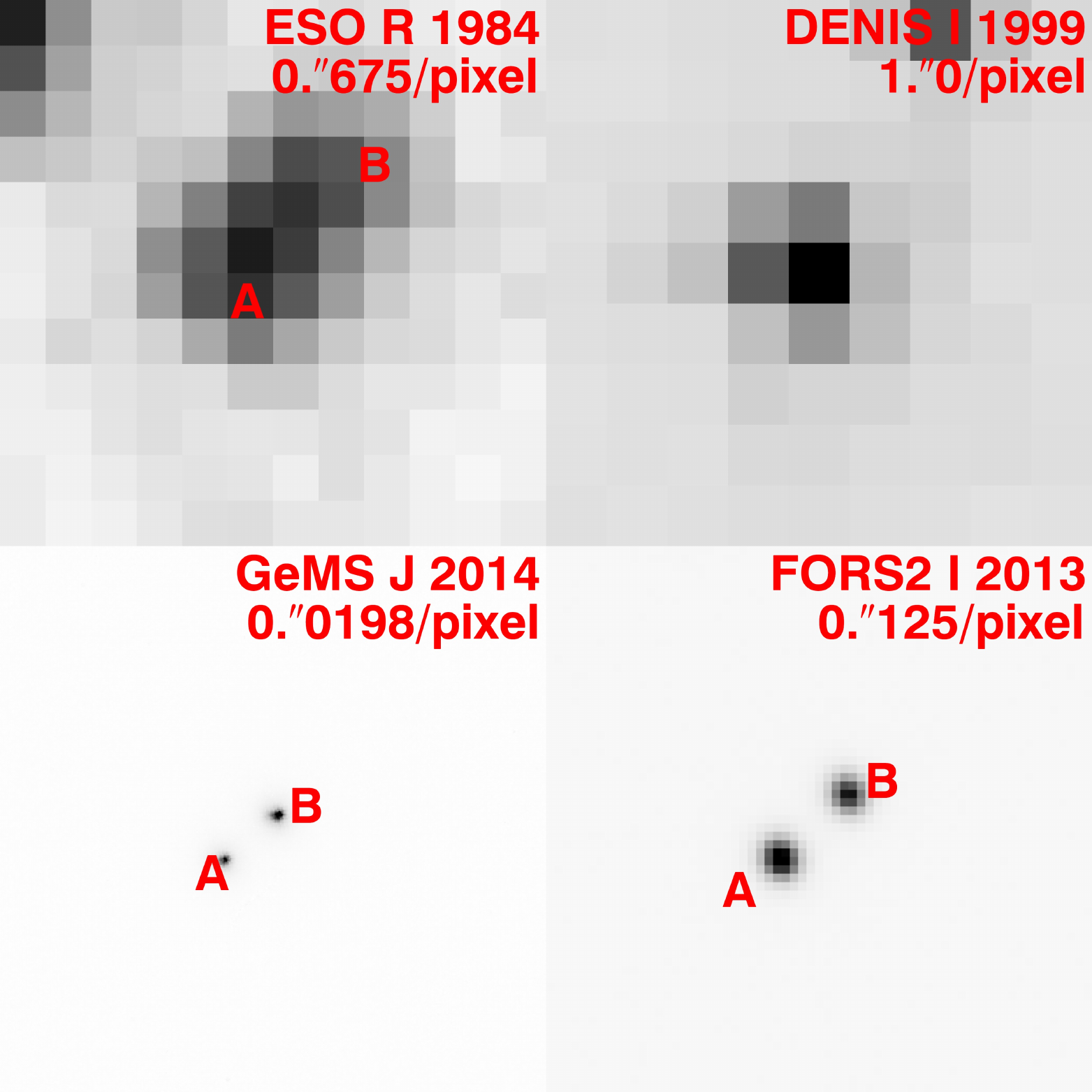}
	\caption{Sample images of Luhman 16 AB used to derive barycenter and orbital astrometry listed in Tables~\ref{table:relastro}~and~\ref{table:absastro}. All images are the same $9^{\prime\prime}\times9^{\prime\prime}$ cut out and rotated north up. Luhman 16 B is brighter than Luhman 16 A for the GeMS images (lower-left panel) due a $J$ band flux reversal across the L-T transition as noted by \cite{Faherty14}. }
	\label{fig:sampdata}
\end{center}
\end{figure}

\begin{figure}[ht!]
\begin{center}
	\includegraphics[width = \textwidth]{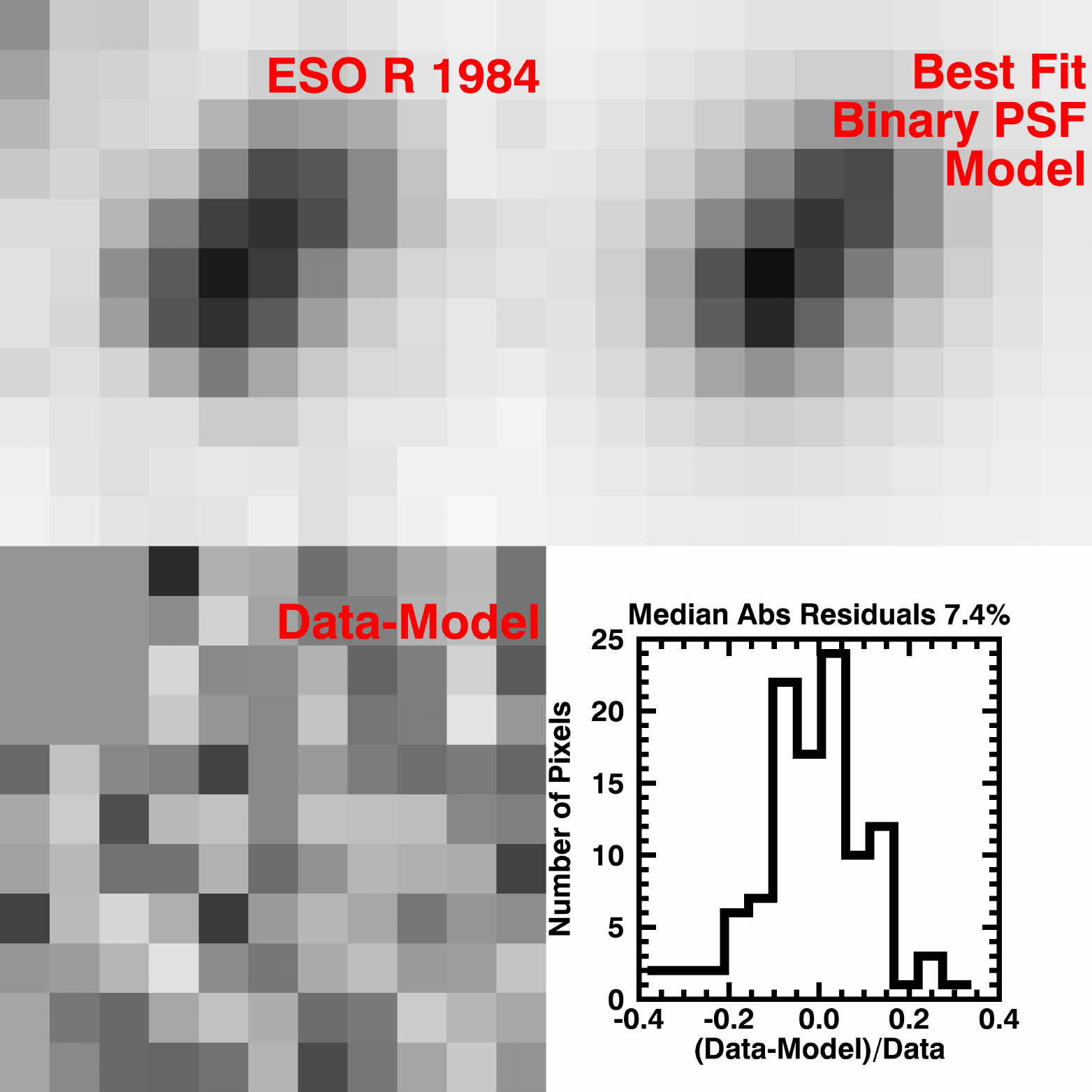}
	\caption{{\bf Top-Left:} ESO 1984 photometric plate image of Luhman 16 AB. {\bf Top-Right:} Minimum-variance binary PSF Lorentizian model (see~\S\ref{sec:eso1984}). {\bf Bottom-Left:} Minimum-variance ``Data-Model" residuals, with the upper right few pixels masked due to the presence of another star. {\bf Bottom-Right:} The histogram of the residuals are Gaussian distributed, with median absolute residuals of $7.4$\%.}
	\label{fig:1984resid}
\end{center}
\end{figure}

\begin{figure}[ht!]
\begin{center}
	\includegraphics[width = \textwidth]{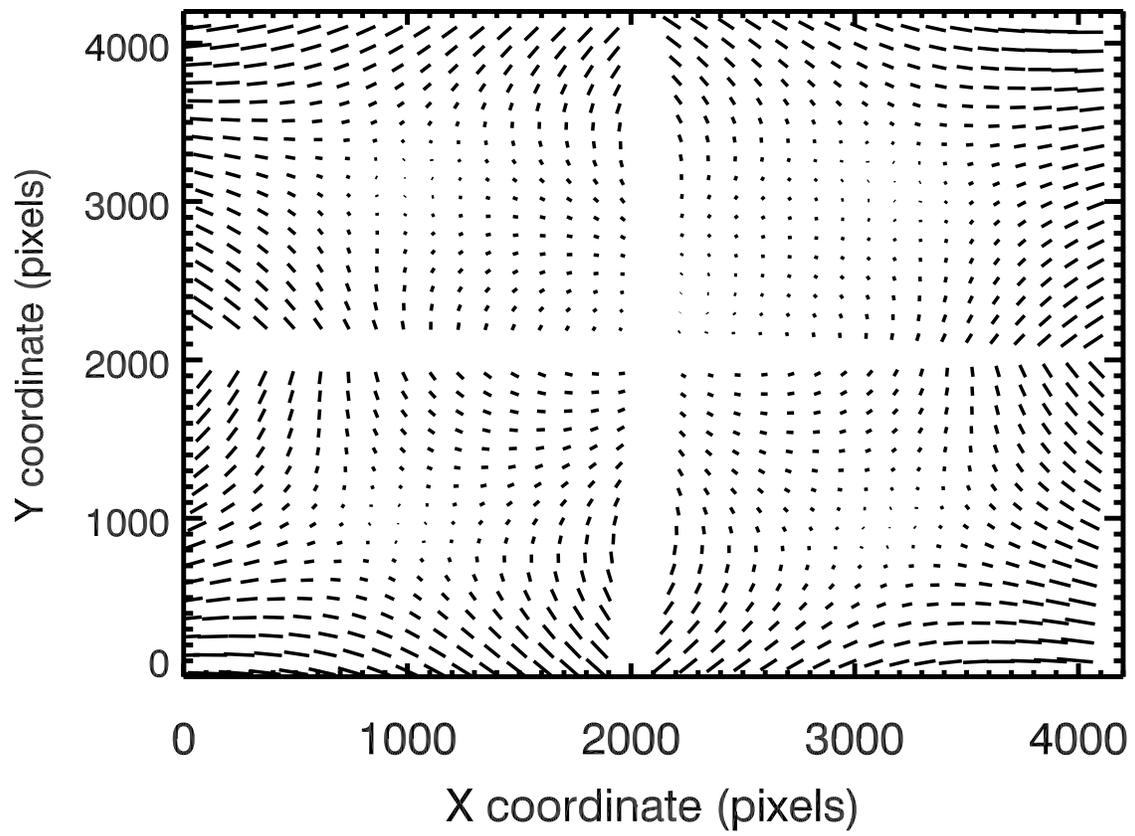}
	\caption{Nonlinear components of GSAOI distortion map.  Plotted vectors and distortion map are derived as described in Section \ref{sect:gems_distortion}.}
	\label{fig:GSAOI_distortion}
\end{center}
\end{figure}

Luhman 16 AB is unresolved in 1999 as shown in Figure~\ref{fig:sampdata} (upper right panel).  We quantify this upper limit on the binary separation at all position angles as follows. First, we use the same procedure detailed in \S\ref{sec:eso1984}, to find a minimum-variance circular Lorentzian PSF model to background stellar PSFs in the image.  We use Lorentzian models rather than Gaussian models as the mean reduced chi-square $\chi^2_{\rm red}$ 
was significantly lower (1.17 vs 1.96). We fit our model Lorentzian PSF to a $7\times7$ pixel (1\farcsec0/pixel$^{-1}$ plate scale) region centered on each stellar PSF, finding an average Lorentzian FWHM of 0\farcsec79$\pm$0\farcsec05 for reference stars in the field nearby Luhman 16 AB.  We used these average Lorentzian parameters as input for the next step.   

\begin{figure}[ht!]
\begin{center}
	\includegraphics[width=3.0 in, height = 3.0 in]{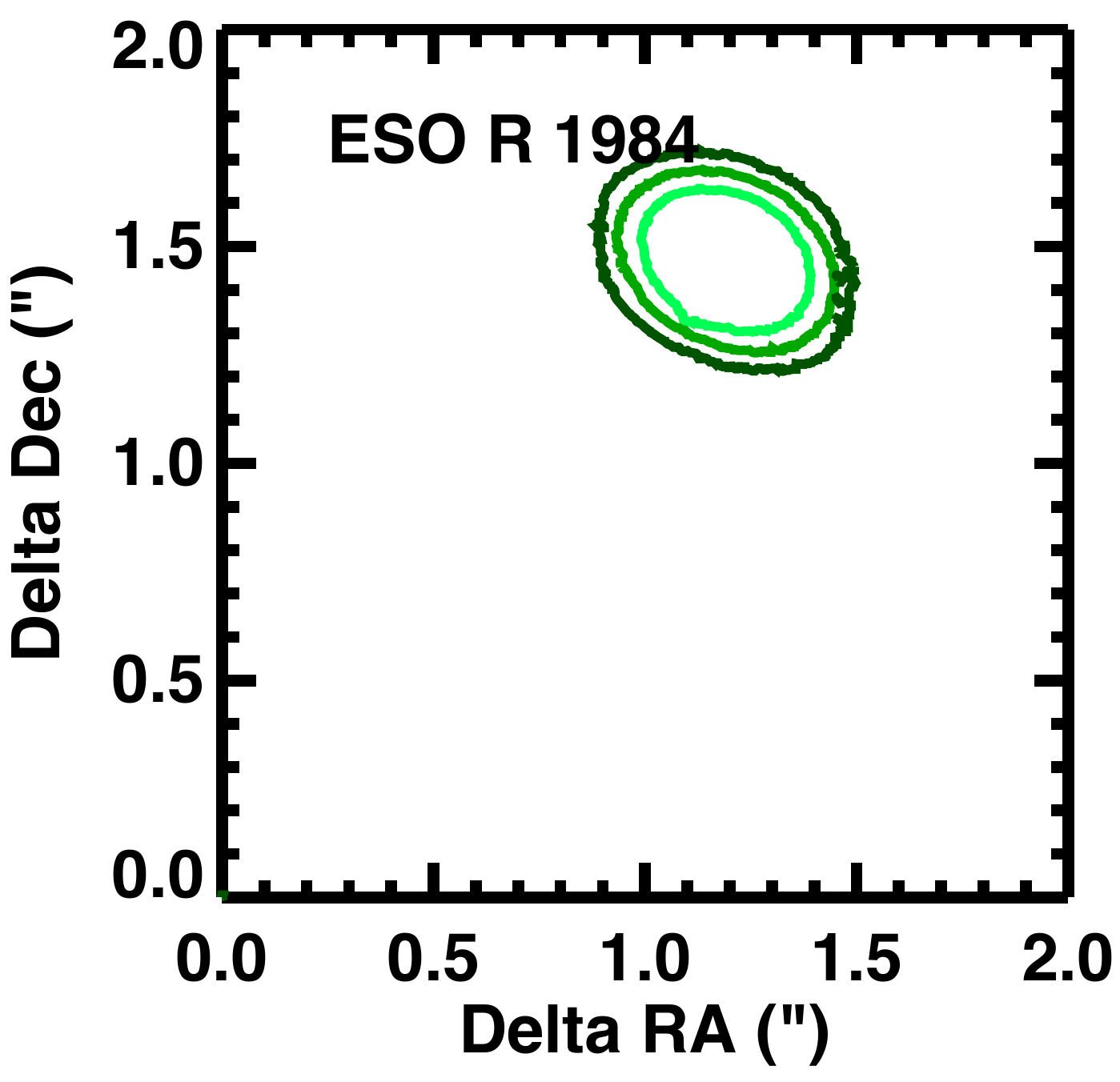}
	\includegraphics[width=3.0 in, height = 3.0 in]{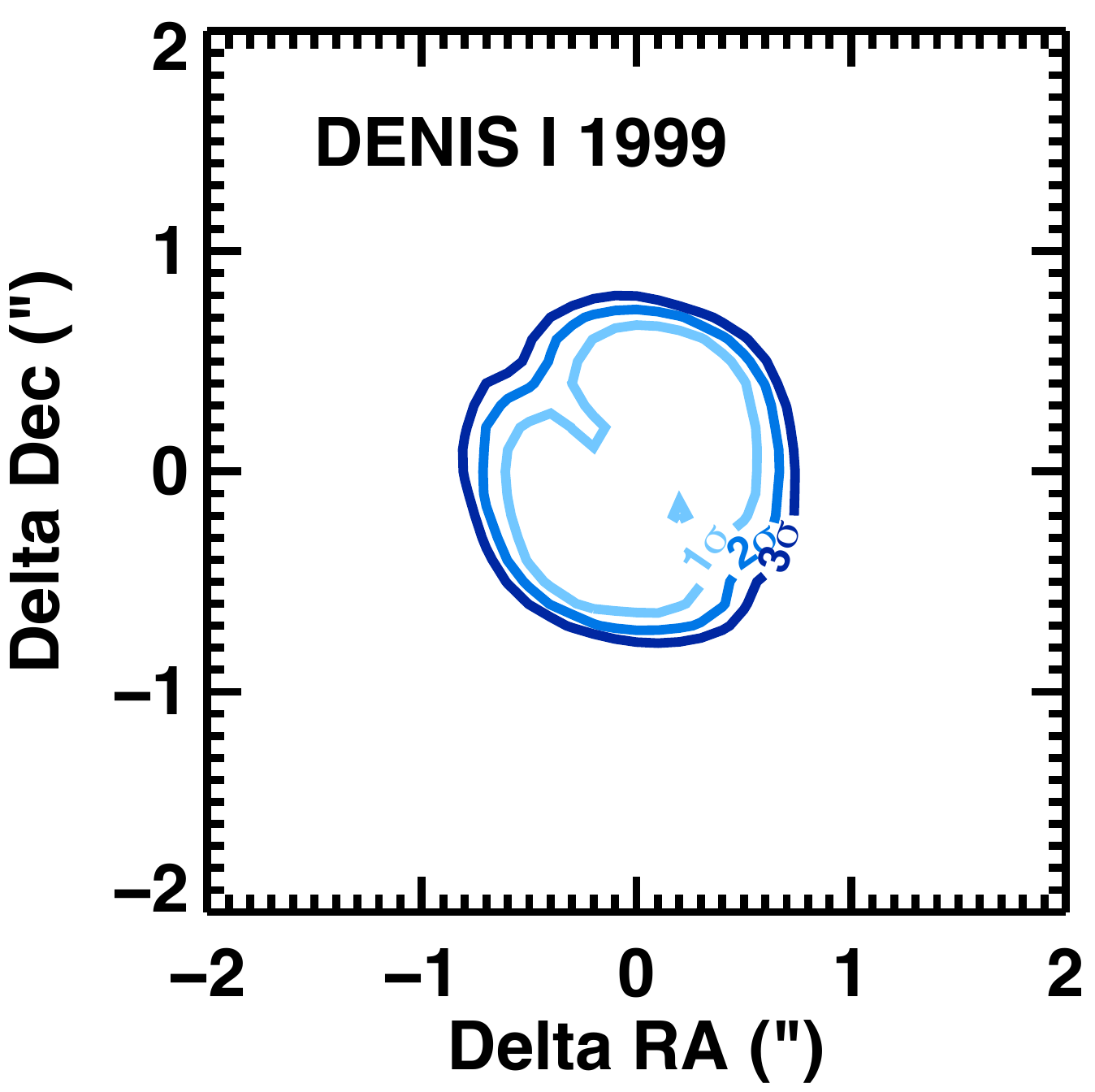}
	\caption{{\bf Left:} $\Delta\chi^{2}$ 1$\sigma$, 2$\sigma$ and 3$\sigma$ contours for the binary separation from our binary Lorentizian PSF model fits to the 1984 ESO resolved photometric plate of Luhman 16 AB (see~\S\ref{sec:eso1984}). We find a minimum-variance separation of~\bestfitsepeso~mas and PA of~\bestfitpaeso, in good agreement with the 138$^{\circ}$ elongation measured from this photometric plate by~\cite{Mamajek13}. {\bf Right:} Contours for the binary separation from DENIS $I$-band unresolved image of Luhman 16 AB (see~\S\ref{sec:denis99}), which places a constraint on a maximum $\Delta $RA~$\lesssim675$ mas and $\Delta$Dec~$\lesssim675$ mas, due to excellent seeing conditions. 
	}
	\label{fig:conts}
\end{center}
\end{figure}

Similar to \S\ref{sec:eso1984}, we use a binary PSF of two co-added, scaled, and shifted Lorentzians to place a constraint on the maximum possible separation of the Luhman 16 AB binary while remaining consistent with the unresolved DENIS image. We minimized the $\chi^{2}$ using \texttt{MPFIT} as above, with five parameters of interest: the position $X_1$,$Y_1$ of the primary, the position $X_2$,$Y_2$ of the secondary, and the total flux of the system. We started our binary-fitting routine with initial guesses from a grid of $40\times40$ different positions, with step sizes of 0\farcsec1, for a total of $1600$ runs. A contour map of $\Delta\chi^{2}=\chi^{2}-\chi^{2}_{\rm min}$, is shown in as the solid dark blue, blue, and light blue lines in Figure~\ref{fig:conts} (right panel), for $1\sigma$, $2\sigma$, and $3\sigma$ confidence. We find the best fit can position angle can vary by $\pm180^{\circ}$ (dotted lines in Figure~\ref{fig:conts}, and we use both solutions in fitting the orbit of Luhman 16 AB (see~\S\ref{sec:orb}).

\subsection{Uncertainties on Stellar Positions \label{sec:abserr}}

There is significant systematic epoch-to-epoch variation in the observations of Luhman 16 AB with GeMS and FORS2. We cannot characterize systematic uncertainty for Luhman 16 using Luhman 16's position measurements directly, given that the binary has orbital motion across multiple epochs. However, we can at least characterize the random terms on a per-epoch basis by computing the standard deviation of the Luhman 16 AB position measurements in a set of frames within an epoch $t$ (we denote this parameter $\sigma_{\alpha_A}(t)$. To estimate the systematic component of the position uncertainty, we measure the epoch-to-epoch variation of the mean position of background stars as described below.  

Subsection \ref{sec:fors2abserr} describes the measurement of position uncertainties in the FORS2 reference frame, subsection \ref{sec:gemsabserr} details the measurement of position uncertainties in the GeMS reference frame, and subsection \ref{sec:seperror} describes the methods used to estimate the uncertainties on the Luhman 16 AB separation.  Later in the paper, subsection \label{sec:baryerror} gives the methodology for estimating the Luhman 16 AB barycenter uncertainties from the position uncertainties.

\subsubsection{Stellar Position Uncertainties for FORS2 \label{sec:fors2abserr}}
We assign uncertainties on the mean position measurements of the individual components of Luhman 16 AB in the FORS2 reference frame at an epoch $t$ as follows:
\begin{equation}
    \begin{split}
        \sigma^{2}_{\rm F, \alpha_A, tot}(t) = \sigma^{2}_{\rm F, \alpha_A}(t) / N(t) + \sigma^{2}_{\rm F, sys, L16}\\
        \sigma^{2}_{\rm F, \delta_A, tot}(t) = \sigma^{2}_{\rm F, \delta_A}(t) / N(t) + \sigma^{2}_{\rm F, sys, L16}\\
        \sigma^{2}_{\rm F, \alpha_B, tot}(t) = \sigma^{2}_{\rm F, \alpha_B}(t) / N(t) + \sigma^{2}_{\rm F, sys, L16}\\
        \sigma^{2}_{\rm F, \delta_B, tot}(t) = \sigma^{2}_{\rm F, \delta_B}(t) / N(t) + \sigma^{2}_{\rm F, sys, L16}
    \end{split}
\end{equation}

where $\sigma_{\rm F, \alpha_{A/B}}(t)$ and $\sigma_{\rm F, \delta_{A/B}}(t)$ are the standard deviations of the FORS2 position measurements of Luhman 16 A/B over all frames within an epoch $t$, $N(t)$ is the number of frames in epoch $t$, and $\sigma_{\rm sys, L16}$ is an estimate of the systematic component of the uncertainty of the positions of the individual components of Luhman 16.

Stars in the background have predictable celestial motion and can serve as ideal references whose position residuals, following subtraction of a fitted proper motion and parallax model, reflect a more accurate measure of the total uncertainty.  We assume that the total uncertainties on the Luhman 16 positions are comparable to the uncertainties of background stars of similar brightness:

\begin{equation}
median(\sigma_{\rm F, tot, L16}(t)) \sim \sigma_{\rm F, std, bkgd}
\end{equation}

\begin{figure}[ht!]
\begin{center}
	\includegraphics[trim={3cm 2cm 3cm 3.8cm},clip,width= 3.2 in]{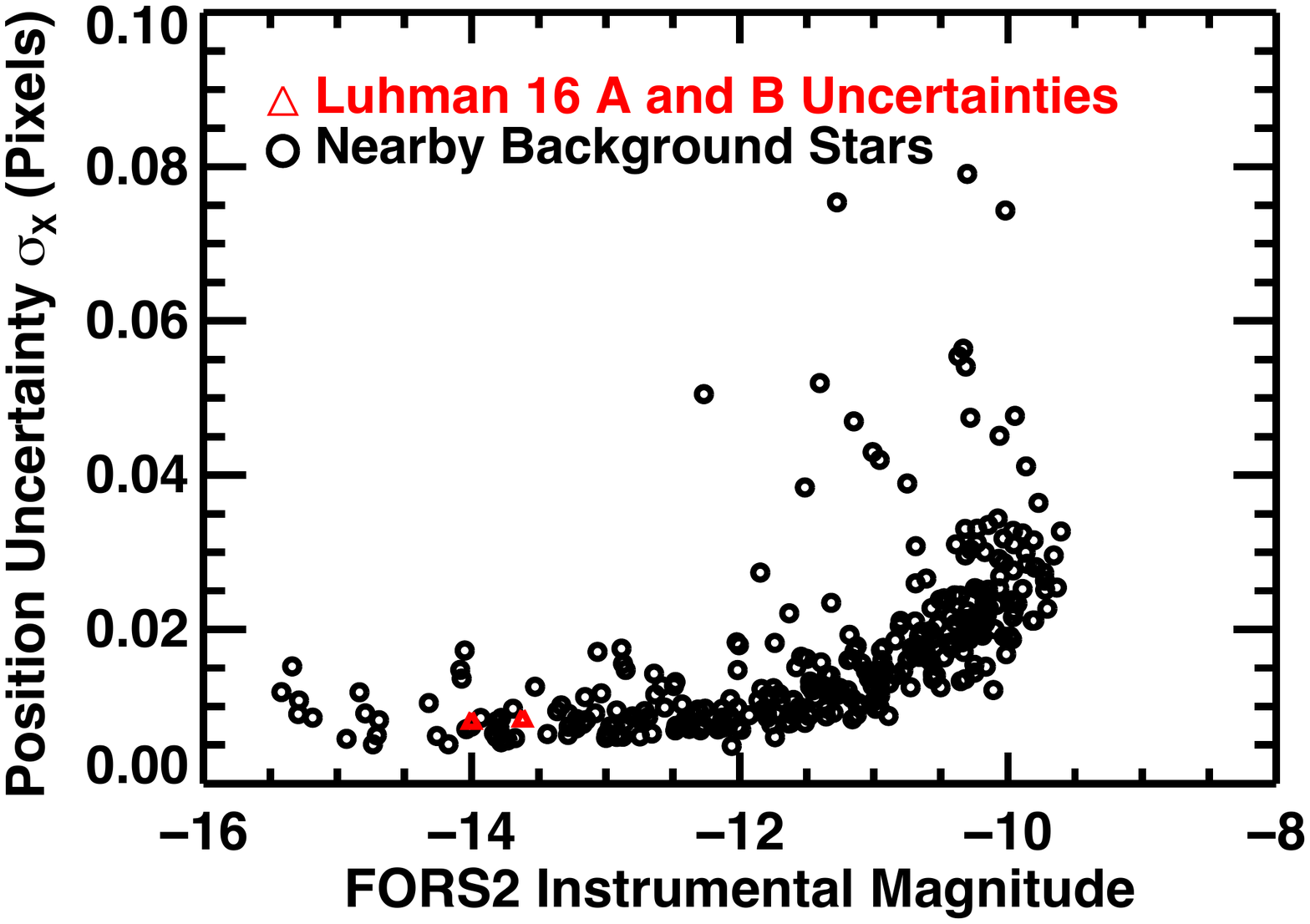}
		\includegraphics[trim={3cm 2cm 3cm 3.8cm},clip,width = 3.2 in]{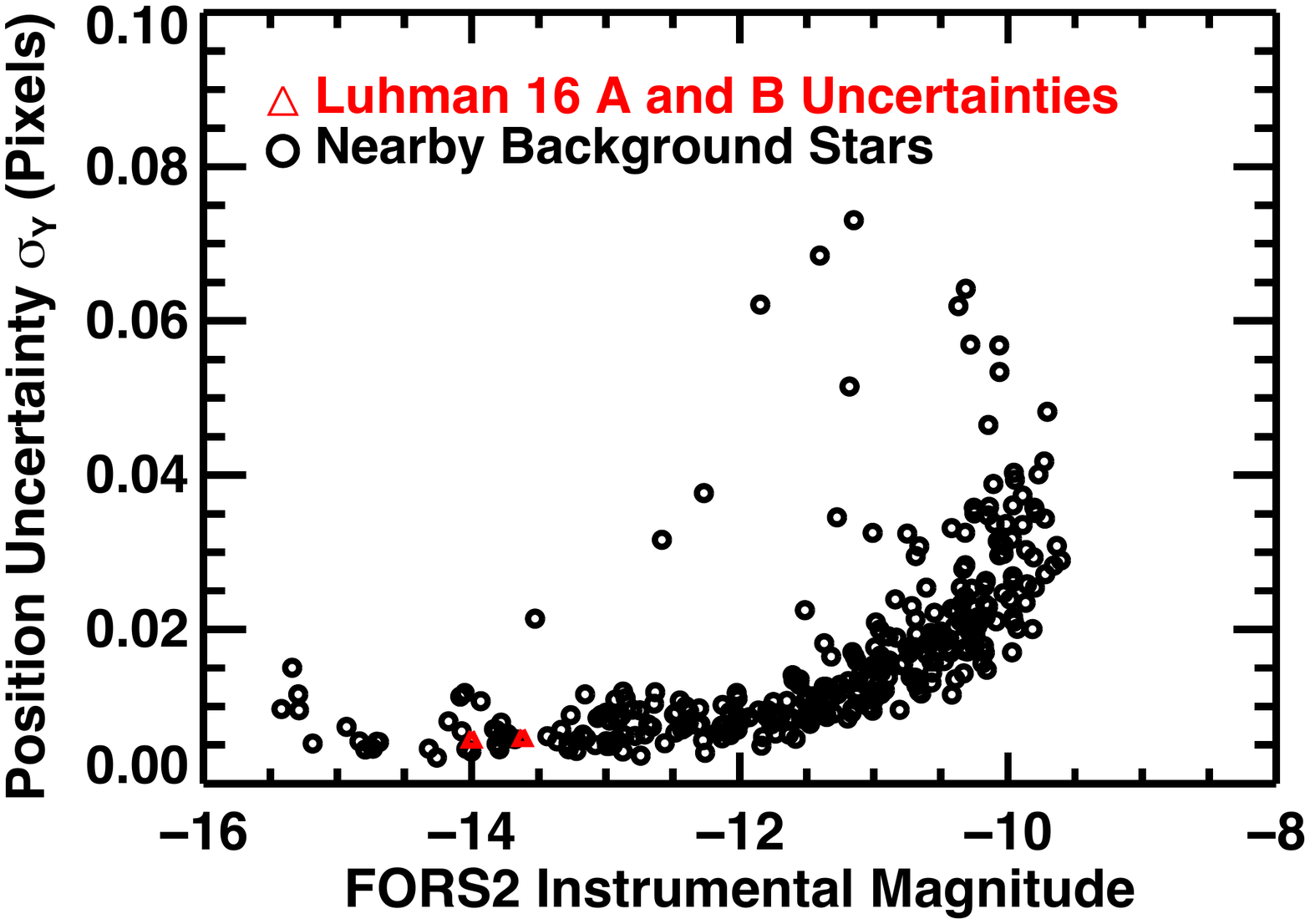}
	\caption{Position uncertainties over all epochs for VLT/FORS2 observations of background stars (black circles) and Luhman 16 AB (red triangles).  The plotted uncertainties for Luhman 16 AB are the median values calculated over all epochs.
	}
	\label{fig:scale}
\end{center}
\end{figure}

To estimate $\sigma_{\rm F, sys, bkgd}$, we first obtain the mean positions of isolated background stars averaged over frames in each epoch $t$. Next, we fit and subtract a minimum-variance parallax/proper motion model for each background star. For the set of background stars nearby Luhman 16 AB in the image and of comparable brightness, we compute the standard deviation of the residual positions over all epochs (black circles, Figure~\ref{fig:scale}).

Substituting into equation 2, the scalar $\sigma_{\rm F, sys, L16}$ can then be approximated as
\begin{equation}
\sigma^{2}_{\rm F, sys, L16} \sim \sigma^{2}_{\rm F, std, bkgd} - median(\sigma^{2}_{\rm F, {\alpha/\delta}_{A/B}}(t) / N(t))
\end{equation}
which is a function of quantities measurable from the data itself.  The scalar $\sigma_{\rm F, sys, L16}$ can then be used in eq. 2 to calculate the vectors of uncertainties $\sigma_{\rm F, {\alpha}_{A/B}, tot}(t)$ and $\sigma_{\rm F, {\delta}_{A/B}, tot}(t)$.  

To estimate the errors in the ICRS reference frame, we add in quadrature a term $\sigma_{\rm F-I}$ accounting for uncertainties introduced by converting the FORS2 astrometric system to the ICRS system.  We estimate $\sigma_{\rm F-I}$ as the standard deviation of the reference star positions in 2015 as measured by GAIA minus the transformed FORS2 positions at the 2015 epoch.  This value, averaged over both axes, is $1.14$ mas. Our final position uncertainties in the ICRS system are 

\begin{equation}
\begin{split}
\sigma^{2}_{\rm I, {\alpha}_{A/B}, tot}(t) = \sigma^{2}_{\rm F, {\alpha}_{A/B}, tot}(t) + \sigma^{2}_{\rm F-I}\\
\sigma^{2}_{\rm I, {\delta}_{A/B}, tot}(t) = \sigma^{2}_{\rm F, {\delta}_{A/B}, tot}(t) + \sigma^{2}_{\rm F-I}
\end{split}
\end{equation}

and are given in Table~\ref{table:absastro}.

\subsubsection{Stellar Position Uncertainties for GeMS \label{sec:gemsabserr}}
Given that we have only 1-4 frames per epoch, we are unable to estimate a per-epoch uncertainty using the methods as above (\S\ref{sec:fors2abserr}). The PSF FWHMs delivered by the GeMS AO system do not vary significantly across epochs ($\sim100$ mas FWHM, $\sim20\%$ variation) and the signal-to-noise ratios of the detections are typically quite high ($500-1000$), so the contribution from random noise sources like shot noise and read noise is expected to be small ($\sim0.1$ mas).  The uncertainties are dominated by terms associated with conversion between reference frames rather than poor signal-to-noise.  Thus, we assign the same Luhman 16 AB position uncertainty to all epochs.

We estimate the position uncertainty for Luhman 16 AB in the ICRS reference frame in both R.A. and Decl. as:
\begin{equation}
\sigma_{\rm I, tot}^2 = \sigma_{\rm F-I}^2 + \sigma_{\rm Ge-F}^2 + \sigma_{\rm GeMS-internal}^2
\end{equation}
where the additional term $\sigma_{\rm Ge-F}=1.22$ mas, as above (\S\ref{sec:fors2abserr}), models uncertainties introducted by transforming the GeMS astrometric system to the FORS2 system.  We estimate the uncertainty $\sigma_{\rm GeMS-internal} = 0.41$ mas by taking the standard deviation of the mean of background star positions over all epochs.  We estimate $\sigma_{\rm Ge-F}$ by computing the standard deviation of the reference star positions as measured by FORS2 minus the transformed GeMS positions.  The sum of all uncertainties in quadrature is $\sigma_{\rm I, tot} = 1.72$ mas.


\subsection{Luhman 16 AB Separation Uncertainties for FORS2 and GeMS\label{sec:seperror}}





The uncertainty on the mean separation between components of Luhman 16 AB is expected to be smaller than the uncertainties on the mean positions of the individual components.  Most sources of uncertainty are correlated for closely-spaced stars, including the Differential Tip/Tilt Jitter (DTTJ), PSF anisotropy, and other systematic terms \citep{Ammons11,Ammons12}. Thus, we make estimates of the uncertainties on the separations $\sigma_{\rm \alpha_A - \alpha_B}(t)$, $\sigma_{\rm \delta_A - \delta_B}(t)$ at epoch $t$ independently from the uncertainties on the mean positions of the individual components as done in \S\ref{sec:abserr}.

In the FORS2 reference frame, the uncertainties on the mean separation measurements of the individual components of Luhman 16 AB at an epoch $t$ are:
\begin{equation}
    \begin{split}
        \sigma^{2}_{\rm F, \alpha_A - \alpha_B, tot}(t) = \sigma^{2}_{\rm F, \alpha_A - \alpha_B}(t) / N(t) + \sigma^{2}_{\rm F, sys, L16}\\
        \sigma^{2}_{\rm F, \delta_A - \delta_B, tot}(t) = \sigma^{2}_{\rm F, \delta_A - \delta_B}(t) / N(t) + \sigma^{2}_{\rm F, sys, L16}\\
    \end{split}
\end{equation}

where $\sigma^{2}_{\rm F, \alpha_A - \alpha_B}(t)$ and $\sigma^{2}_{\rm F, \delta_A - \delta_B}(t)$ are the standard deviations of the separation measurements over frames within an epoch $t$.  We estimate the systematic component of the uncertainty on the separation of Luhman 16 AB $\sigma_{F, sys, L16}$ using a calibration background double star. The background double star has a separation of $1\farcs8$ and a position of 10:49:11.81 -53:18:57. This calibration star is $\approx0\farcmin4$ away from Luhman 16 AB. We chose this double star because it has a similar separation and flux ratio as the components of Luhman 16 AB. However, the double star is fainter by $\approx2$ mag in $I$-band and thus has more poisson noise.  We assume that the true space motion of this double star is defined by purely parallax and proper motion for the two components.  With a component magnitude of $V\approx18.5$, the double star is likely composed of unbound stars at distances beyond 1 kpc. 

For the GeMS data, the double star used above does not yield a usable calibration of the uncertainty as the S/N is lower. There are only between 1 and 3 frames per epoch in the Luhman 16 separation data from GeMS, so there is not enough data to estimate the uncertainties from the data itself for each epoch.  Instead, for epochs with more than 1 frame, we conservatively compute the standard deviation of the separation measurements, and average these over all epochs to estimate a single uncertainty on the separation measurement, $0.26$ mas for RA and $0.28$ mas for Declination.
 

\section{MCMC Analysis \label{sec:mcmcall}} 

We constrain a joint 16 parameter astrometry model using a Markov Chain Monte Carlo (MCMC) searching algorithm. This model includes the orbital astrometry parameters (Keplerian orbital parameters and total dynamical mass) as well as the barycentric astrometry parameters (mass ratio, parallax, barycentric proper motion, position at a reference epoch, and CDAR parameters for both FORS2 and GeMS data sets). The starting priors for each parameter are given in Table~\ref{table:paramrange}.  With our simultaneous fit of all parameters, we can inspect joint posterior probability distributions for correlated uncertainty between model parameters derived from orbital and barycentric astrometry.  We present the results of this analysis as well as the individual masses $M_A$ and $M_B$ for Luhman 16 A and B in Section \ref{sect:results}.

We perform the following procedure in every step of our MCMC chains, detailed in the next few sections:  
\begin{enumerate}
    \item Subtract the chromatic differential atmospheric refraction from the coordinate positions for both brown dwarfs (see \S\ref{sec:cdarmod}). 
    \item Compute the Luhman 16 separation for all data from the CDAR-corrected coordinate positions, and then compute the logarithm of the likelihood (``log likelihood'') for the relative orbit terms $L_{\rm rel}$ (see \S\ref{sec:orb}). 
    \item Compute relative radial velocity log likelihood $L_{\rm RV}$ (see~\S\ref{sec:relrv}).
    \item Compute the barycenter positions and the orbit log likelihood $L_{\rm abs}$ from the CDAR-corrected coordinate positions for ESO 1984, FORS2, and GeMS data (see \S\ref{sec:abs}). 
    \item Compute the sum of the negative log likelihood for the orbital,  barycentric motion, radial velocity parameters $L_{\rm rel}+L_{\rm abs}+L_{\rm RV}$. 
\end{enumerate}
We use uniformly distributed priors for all 16 parameters, with ranges shown in Table~\ref{table:paramrange}. The maximum-likelihood parameters and 68.3\% and 95.4\% confidence intervals from our MCMC analysis are shown in Table \ref{table:bestfit}. 

\subsection{Modeling Chromatic Differential Atmospheric Refraction\label{sec:cdarmod}}
    
Chromatic Differential Atmospheric Refraction (CDAR) produces a change in the measured separation vector between any two stars that depends on the zenith angle, the parallactic angle, atmospheric pressure, atmospheric temperature, and the two stellar spectra.  Astrometric measurements in VLT/FORS2 has the additional complication that a majority of the CDAR offset is corrected with the Longitudinal Atmospheric Dispersion Corrector (LADC), a maneuvering prism with a single degree of freedom \citep{avila97}.  

At each step in our MCMC chains, we model the CDAR for both the FORS2 and GeMS astrometry following \citet{lazorenko09}, \citet{sahlmann13}, and \citet{Sahlmann15}. In measuring proper motion and parallaxes of single stars in FORS2 data, these authors find that the offsets produced by CDAR from the atmosphere and the LADC are sufficiently modeled by two parameters $\rho$ and $d$, in units of mas. The first of these, $\rho$, accounts for the atmospheric CDAR offset and the second, $d$, accounts for the LADC correction of CDAR. The correction of CDAR by the LADC is imperfect because the LADC corrects for CDAR at the zenith angle of the beginning of a set of observations, but does not update during the observation. There is a small offset between the actual CDAR $\rho$ and the correction of CDAR by the LADC $d$.

Thus, we include the $\rho$ for FORS2, $\rho_{\rm FORS2}$; the $d$ for FORS2, $d_{\rm FORS2}$; and the $\rho$ for GeMS, $\rho_{\rm GeMS}$ in our MCMC analysis. The prior constraints on these parameters are shown in Table~\ref{table:paramrange}. We place a prior on $\rho_{\rm FORS2}$ and $d_{\rm FORS2}$ to be opposite signs, choosing $(p > 0, d < 0)$, given that we define the atmospheric CDAR offset to always by positive, and the correction to always be negative. Similarly, we place a positive prior constraint on $\rho_{\rm GeMS} > 0$. Finally, we constrain both $\rho_{\rm FORS2}$ and $d_{\rm FORS2}$ to be within reasonable physical ranges for total possible atmospheric CDAR and its correction. 

Both $\rho$ and $d$ include dependencies on the spectrum of the target star.  For the case of FORS2 imaging of Luhman 16 AB, the individual stellar components have different spectra, and a separate $\rho$ and $d$ would have to be fitted for each star individually.  Rather than add these new terms to the global model fit, we explicitly measure the color index $CI_{\rm A,B}$ of each star using MagE medium-resolution ($\sfrac{\lambda}{\Delta\lambda}\approx4000$) spectra and FIRE near-infrared ($\sfrac{\lambda}{\Delta\lambda}\approx8000$) spectra \citep[their Figure 1]{Faherty14}. 


We write the CDAR parameters for Luhman B as a function of the $\rho$ and $d$ for Luhman A and the ratio of color indices $CI_{\rm A}/CI_{\rm B}$. The CDAR parameters $\rho$ and $d$ are expressed in units of mas~$(\tan{z})^{-1}$, where $z$ is the zenith angle of the observation. We compute $\rho$ and $d$ for both Luhman 16 A and B as:
\begin{equation}
\begin{split}
\rho_{\rm A} = \rho \\
\rho_{\rm B} = \rho(CI_{\rm B}/CI_{\rm A})  \\
d_{\rm A} = d \\
d_{\rm B} = \rho(CI_{\rm B}/CI_{\rm A}) 
\end{split}
\end{equation}
The color indices $CI_{\rm AB}$ are defined as the shift in the position of the stellar photocenter due to the effect of CDAR.  We model this using formula (6) from \citet{Gubler98} as follows:

\begin{equation}\label{eqn:ci}
CI = \frac{\int_{\lambda_{min}}^{\lambda_{max}} A(\lambda)T(\lambda)I(\lambda) d\lambda}{\int_{\lambda_{min}}^{\lambda_{max}} T(\lambda)I(\lambda) d\lambda}
\end{equation}
where $T(\lambda)$ is the wavelength-dependent throughput of the atmosphere, telescope, and filters, $I(\lambda)$ is the spectrum of the star, and $A(\lambda)$ is the chromatic component of the refraction equation that is linear with the tangent of the zenith angle.  As described in \citet{Gubler98}, the nonlinear component of the refraction $B(\lambda)$ is three orders of magnitude smaller than the linear component and negligible for the zenith angles of concern here.  We use the function \texttt{refco} within the python implementation of \texttt{SLALIB} to calculate $A(\lambda)$ for observatory parameters corresponding to the VLT, for an altitude of 2635 meters, pressure of 800 mbar, temperature of 278~K, and relative humidity of 10\%.  Stellar spectra $I(\lambda)$ in the $I$-band are obtained from \citet{Faherty14}.  $T(\lambda)$ is calculated as the product of the $I$-band filter function, the QE curve of the detectors in FORS2, and the atmospheric transmission\footnote{We use the Cerro Pachon model with 1.5 air masses and a water vapor column of 4.3 mm.\\ 
See here: http://www.gemini.edu/sciops/telescopes-and-sites/observing-condition-constraints/ir-transmission-spectra}. The ratio in the color indices between Luhman 16 A and B in FORS2 passband is $(CI_{\rm B}/CI_{\rm A}) = 0.985763$ mas~$(\tan{z})^{-1}$.  

Similarly, we calculated the color indices for GeMS observations using the $J$-continuum narrow band filter using Equation~\ref{eqn:ci}, detector characteristics corresponding to GSAOI, atmospheric parameters appropriate for the Gemini South Observatory. Due to the similarity of the $J$-continuum spectra of Luhman 16 A and B as measured in \citet{Faherty14} and the narrow width of the filter band, the ratio in color indices for the GeMS observations evaluates to $0.9999$ mas~$(\tan{z})^{-1}$.  We neglect this correction as it is significantly smaller than the error bars of the GeMS observations.

Following \citet{lazorenko09}, \citet{sahlmann13}, and \citet{Sahlmann15}, we express the CDAR correction in right ascension and declination as a function of parameters $f_{\rm 1,x,m}$, $f_{\rm 1,y,m}$, $f_{\rm 2,x,m}$, and $f_{\rm 2,y,m}$. We use equations 2 and 3 of \cite{sahlmann13} to compute $f_{\rm 1,x,m}$, $f_{\rm 1,y,m}$, $f_{\rm 2,x,m}$, and $f_{\rm 2,y,m}$ using parameters related to zenith angle, atmospheric pressure, atmospheric temperature, and parallactic angle from Table~\ref{table:cdar}. We re-compute and then subtract off CDAR from the coordinate positions at each step of our MCMC chains by fitting for $\rho$ and $d$ for both the FORS2 and GeMS data:
    \begin{equation}\label{eqn:cdarfors2}
    \begin{split}
    \alpha^{\prime}_{A,\rm FORS2} = \alpha_{A,\rm FORS2}+(\rho_{\rm A,\rm FORS2}f_{\rm 1,x,m}+d_{\rm A,\rm FORS2}f_{\rm 2,x,m})\\
    \delta^{\prime}_{A,\rm FORS2} = \delta_{A,\rm FORS2}-(\rho_{\rm A,\rm FORS2}f_{\rm 1,y,m}+d_{\rm A,\rm FORS2}f_{\rm 2,y,m})\\
    \alpha^{\prime}_{B,\rm FORS2} = \alpha_{B,\rm FORS2}+(\rho_{\rm B,\rm FORS2}f_{\rm 1,x,m}+d_{\rm B,\rm FORS2}f_{\rm 2,x,m})\\
    \delta^{\prime}_{B,\rm FORS2} = \delta_{B,\rm FORS2}-(\rho_{\rm B,\rm FORS2}f_{\rm 1,y,m}+d_{\rm B,\rm FORS2}f_{\rm 2,y,m}) \\
    \end{split}
    \end{equation}
where $(\alpha^{\prime}_{AB,\rm FORS2}, \delta^{\prime}_{AB,\rm FORS2})$ are the CDAR-corrected coordinate positions for a given MCMC step and given $\rho_{\rm AB,\rm FORS2}$,$d_{\rm AB,\rm FORS2}$ parameters, and  $(\alpha_{AB,\rm FORS2}, \delta_{AB,\rm FORS2})$ is the uncorrected FORS2 coordinate positions. For the GeMS data, as mentioned above, we neglect the difference in color indices between Luhman 16 AB and use a single $\rho_{\rm GeMS}$ parameter to model CDAR:
    \begin{equation}\label{eqn:cdargems}
    \begin{split}
    \alpha^{\prime}_{A,\rm GeMS} = \alpha_{A,\rm GeMS}+(\rho_{\rm GeMS} f_{1,x,m})\\
    \delta^{\prime}_{A,\rm GeMS} = \delta_{A,\rm GeMS}-(\rho_{\rm GeMS} f_{1,y,m})\\
    \alpha^{\prime}_{B,\rm GeMS} = \alpha_{B,\rm GeMS}+(\rho_{\rm GeMS} f_{1,x,m})\\
    \delta^{\prime}_{B,\rm GeMS} = \delta_{B,\rm GeMS}-(\rho_{\rm GeMS} f_{1,y,m})
    \end{split}
    \end{equation}
where $(\alpha^{\prime}_{AB,\rm GeMS}, \delta^{\prime}_{AB,\rm GeMS})$ are the CDAR corrected positions for a given MCMC step and a given $\rho_{\rm GeMS}$, and  $(\alpha_{AB,\rm GeMS}, \delta_{AB,\rm GeMS})$ are the uncorrected GeMS positions. We do not subtract the effects of CDAR at the 1984 and 1999 epochs, which are negligible relative to the substantial astrometric uncertainty for those data. 

The photocenter shift due to CDAR expected for the 1984 ESO Schmitt observations $<50$ mas, which is $4\times$ smaller than the uncertainties on the barycenter position due to other sources of error.  Therefore, we neglect the CDAR correction in the 1984 barycenter position as well as the correction on the relative separation.  For the unresolved 1999 DENIS data, we do not attempt to constrain the barycenter position, and the change in the relative separation due to CDAR is far smaller than the error bars, so we again neglect CDAR.  

\subsection{Orbital Astrometry Model\label{sec:orb}}

At each step in our MCMC chains, we compute the log likelihood of our orbital astrometry model parameters $L_{\rm rel}$.  As the CDAR-subtracted Luhman 16 separations are themselves dependent on the three CDAR parameters, we re-compute the CDAR magnitudes and thus the corrected separations at each step in our MCMC chains:
    \begin{equation}\label{eqn:rel}
    \begin{split}
    \Delta\alpha(t) = (\alpha_{B}^{\prime}(t)-\alpha_{A}^{\prime}(t))\cos{\frac{\delta_{B}^{\prime}(t)+\delta_{A}^{\prime}(t)}{2}} \\
    \Delta\delta(t) = \delta_{B}^{\prime}(t)-\delta_{A}^{\prime}(t)
    \end{split}
    \end{equation}
where $\alpha_{A,B}^{\prime}(t)$,$\delta_{A,B}^{\prime}(t)$ are the CDAR-subtracted coordinates of Luhman 16 AB from Equations~\ref{eqn:cdarfors2}~and~\ref{eqn:cdargems} at epoch $t$ and $\Delta\alpha(t)$,$\Delta\delta(t)$ are the relative separations. Uncertainties on the separations $\sigma_{\Delta\alpha}$,$\sigma_{\Delta\delta}$ for GeMS and FORS2 wide-field imaging are computed as detailed in \S\ref{sec:relgems} and \S\ref{sec:fors2} respectively. 

We model orbital astrometry from Equation~\ref{eqn:rel} with 6 Keplerian orbit parameters: semi-major axis $a$, eccentricity $e$, inclination $i$, longitude of ascending node $\Omega$, argument of periastron $\omega$, and time of periapsis $\tau$ \citep{Green85}. For our MCMC analysis, we adopt uniform priors in total mass $M_{\rm tot} = M_A+M_B$, semi-major axis log~$a$, eccentricity $e$, longitude of ascending node~$\Omega$, argument of periastron $\omega$, time of periastron passage $\tau$ and inclination $\cos{i}$, as shown in Table~\ref{table:paramrange}. 

For the 1999 DENIS data, we explicitly compute probability density functions (PDF) as a function of binary separation and position angle as described in \S\ref{sec:denis99}.  The confidence intervals derived from these PDFs are shown in Figure~\ref{fig:conts}. At each iteration of the MCMC run, the orbital model predicts the separation and P.A. at the 1999 epoch.  We add the log PDF (log 1-$p$) at the model separation and P.A. to the overall likelihood function $L$.  
\subsection{Relative Radial Velocity Model \label{sec:relrv}}

VLT/CRIRES relative radial velocities for Luhman 16 AB (see~\S\ref{sec:crires}) provide complementary information about the mutual orbit to our orbital astrometry above. At each step in our MCMC chains, we compute the relative radial velocity model $V_{A} - V_{B}$ at each of 3 epochs using Equation 2.44, pg 42 from \cite{Hilditch}. We add the likelihood function, $L_{\rm RV}$ to the overall likelihood $L$.  Based on the evolution in the model posteriors before and after the addition of the radial velocity data, we assess that this data is primarily constraining the longitude of ascending node $\Omega$, the argument of periastron $\omega$, and the eccentricity $e$. These constraints disfavored sections of the mutual orbit parameter space that were not ruled out with the orbital astrometry alone. 

\subsection{Barycentric Motion Model\label{sec:abs}}
Luhman 16 AB was unresolved by DENIS in 1999, and the barycenter would be difficult to derive from the photocenter.  Thus we exclude the DENIS $I$-band 1999 epoch from our barycentric motion model. At each step in our MCMC chains, we compute the log likelihood of the barycenter model parameters $L_{\rm abs}$. We compute the position of the barycenter \citep{Sahlmann15} at each epoch $t$ from the CDAR-subtracted coordinate positions (i.e. relative to background stars): 
\begin{equation}\label{eqn:bary}
\begin{split}
\alpha_{\gamma}(t)=\bigg(\frac{1}{1+q}\bigg)(\alpha^{\prime}_A+q\alpha^{\prime}_B)\\
\delta_{\gamma}(t)=\bigg(\frac{1}{1+q}\bigg)(\delta^{\prime}_A+q\delta^{\prime}_B)
\end{split}
\end{equation}  
where $q=\frac{M_B}{M_A}$ is the mass ratio, $\alpha_{\gamma}(t)$,$\delta_{\gamma}(t)$ are the Right Ascension and Declination of the barycenter at epoch $t$, and $\alpha^{\prime}_{A,B}$,$\delta^{\prime}_{A,B}$ are the Luhman 16 AB positions from Equations~\ref{eqn:cdarfors2}~and~\ref{eqn:cdargems}. Our model for the position of the barycenter at epoch $t$ is given as:
\begin{equation}\label{eqn:para}
\begin{split}
\alpha_{\gamma}(t)=\alpha_{0}+\mu_{\alpha}t+\varpi\Pi_{\alpha} \\
\delta_{\gamma}(t)=\delta_{0}+\mu_{\delta}t+\varpi\Pi_{\delta}
\end{split}
\end{equation}
where $\alpha_{\gamma}(t)$,$\delta_{\gamma}(t)$ are the coordinates of the barycenter, $\varpi_{\rm rel}$ is the parallax factor, and $\mu_{\alpha}$,$\mu_{\delta}$ are the proper motions in Right Ascension and Declination.  To maximize $L_{\rm abs}$, we minimize the residuals between Equation~\ref{eqn:bary} and~\ref{eqn:para}. Thus, from Equations~\ref{eqn:bary}~and~\ref{eqn:para}, we model 6 parameters associated with the location of the barycenter - and importantly, the barycenter position changes significantly as a function of mass ratio. For our MCMC analysis, we adopt uniform priors for the reference epoch positions $\alpha_{0}$,$\delta_{0}$, proper motion $\mu_{\alpha}$ and $\mu_{\delta}$, parallax $\varpi_{\rm rel}$, and mass ratio $q$ as shown in Table~\ref{table:paramrange}.

\subsubsection{Derivation of Barycentric Position Errors\label{sec:baryerror}}
\newcommand*\mean[1]{\bar{#1}}
Unlike the relative separation uncertainties, the uncertainties on the barycenter position have a dependence on the mass ratio $q$. Propagated uncertainties on the barycenter $\sigma_{\alpha_{\gamma}}(t)$,$\sigma_{\delta_{\gamma}}(t)$ are computed for the FORS2, GeMS and ESO 1984 plate as: 
\begin{equation}\label{eqn:baryerr}
\begin{split}
\sigma^{2}_{\alpha_{\gamma}}(t) = \bigg(\frac{1}{1+q}\bigg)^{2}\sigma_{\alpha_A}(t)^{2}+\bigg(\frac{q}{1+q}\bigg)^{2}\sigma_{\alpha_B}(t)^{2}+2\frac{q}{1+q}\frac{1}{1+q}\sigma_{\alpha_A,\alpha_B}(t) \\
\sigma^{2}_{\delta_{\gamma}}(t) = \bigg(\frac{1}{1+q}\bigg)^{2}\sigma_{\delta_A}(t)^{2}+\bigg(\frac{q}{1+q}\bigg)^{2}\sigma_{\delta_B}(t)^{2}+2\frac{q}{1+q}\frac{1}{1+q}\sigma_{\delta_A,\delta_B}(t)      
\end{split}
\end{equation}
where $\sigma_{\alpha_A}(t),\sigma_{\alpha_B}(t)$ are the position uncertainties for Luhman 16 AB in the ICRS system at epoch $t$ as derived in \S\ref{sec:abserr}, $q=\frac{M_B}{M_A}$ is the mass ratio, and $\sigma_{\alpha_A,\alpha_B}(t),\sigma_{\delta_A,\delta_B}(t)$ are the correlated error between the positions of Luhman 16 AB. Given that we have only 1$-$3 frames per epoch for our GeMS astrometry, we do not include correlated error in the GeMS barycenter uncertainty calculation. 

We include the correlated error in our computation of the FORS2 barycenter uncertainty (Equation~\ref{eqn:baryerr}). The correlated errors are calculated for each epoch from the frame-by-frame positions of both stars:
\begin{equation}
\begin{split}
    \sigma_{X_A,X_B}(t) = \frac{1}{1-N}\sum_{i=1}^{i=N}\bigg[(X_{A,i}-\mean{X_A})(X_{B,i}-\mean{X_B})\bigg] \\
    \sigma_{\alpha_A,\alpha_B}(t) = 125\sigma_{X_A,X_B}(t) \\
    \sigma_{Y_A,Y_B}(t) = \frac{1}{1-N}\sum_{i=1}^{i=N}\bigg[(Y_{A,i}-\mean{Y_A})(Y_{B,i}-\mean{Y_B})\bigg] \\
    \sigma_{\delta_A,\delta_B}(t) = 125 \sigma_{Y_A,Y_B}(t) \\
\end{split}
\end{equation}
where $\sigma_{X_A,X_B}(t)$,$\sigma_{Y_A,Y_B}(t)$ are the correlated errors in pixels for epoch $t$, using $N$ frames. $X_{A,i}$,$Y_{A,i}$ are the positions of Luhman 16 A at frame $i$ in units of pixels, and $\mean{X_A}$ is the mean position over $N$ frames. We convert the correlated errors $\sigma_{X_A,X_B}(t)$,$\sigma_{Y_A,Y_B}(t)$ pixels to $\sigma_{\alpha_A,\alpha_B}(t),\sigma_{\delta_A,\delta_B}(t)$ arcseconds using the FORS2 pixel scale of $125$~mas~pixel$^{-1}$.


\subsection{Running the MCMC \label{sec:mcmc}}
To determine the posterior probability distributions for the orbit and barycenter astrometry parameters, we performed a parallel-tempered MCMC analysis \citep{Earl05} to maximize the likelihood function $L=L_{\rm rel}+L_{\rm abs}+L_{\rm RV}$ of the model given the data \citep{Gregory11,Kalas13}. We compute the CDAR-subtracted coordinate positions as detailed in \S\ref{sec:cdarmod}. We compute the likelihood for the orbit parameters $L_{\rm rel}$ as described in \S\ref{sec:orb}, the likelihood for barycentric orbit parameters $L_{\rm abs}$ as described in \S\ref{sec:abs}, and the likelihood for the relative radial velocity data as described in \S\ref{sec:relrv}. 

We minimize the negative log likelihood function using the Python implementation of the affine-invariant parallel-tempered ensemble sampler \texttt{PTsampler} from the python module \texttt{emcee v2.1.0}\footnote{http://dan.iel.fm/emcee/current/} \citep{ForemanMackey13}. We use~\nwalkers~walkers and~\nsteps~steps, discarding the first 20\% of steps as the burn-in for each walker. We set temperature $T=$\temp~to maximize the search of parameter space for secondary minima.  We visually inspect the highest temperature chain, finding that the walkers cover the entirety of the prior uniformly-distributed parameter ranges in Table~\ref{table:paramrange}. This means the convergence of our MCMC analysis to the best fit parameters and confidence intervals in Table~\ref{table:bestfit} are not the result of a local minima. We also visually inspect the lowest temperature chain, finding the walkers to stabilize over the the~\nsteps~steps, which confirms the MCMC run convergence to our maximum-likelihood parameters. The Gelman-Rubin statistics, which checks for convergence of MCMC chains run in parallel, were $\approx1.0004$ for all 16 parameters, indicating successful convergence \citep{Gelman92}.

\section{Results \label{sect:results}}

The posterior probability distributions for the individual masses $M_A$,$M_B$, mass ratio $q\equiv\frac{M_B}{M_A}$ and sum of masses $M_A+M_B$ are shown in Figure~\ref{fig:trianglemass}. The posterior probability distributions are smooth, indicating no multiple local minima. As shown in Figures~\ref{fig:trianglerel},~\ref{fig:triangleabs} and~\ref{fig:trianglecross}, we compute posterior probability distributions of each of the 16 model parameters by marginalizing over the other 15 parameters from the zero temperature walker distributions. Our 1$\sigma$ (68.3\% confidence) and $2\sigma$ (95.4\%) credible intervals, derived from our MCMC posterior distributions, are shown in Table~\ref{table:bestfit}. Our individual masses derived for Luhman 16 AB are~\masspri~$M_{J}$ and~\masssec~$M_{J}$~respectively. 

As shown in Figure~\ref{fig:trianglerel}, we find slightly elliptical 2D posterior PDFs, when plotting orbit parameters $\Omega$ vs $a$, $e$ vs $i$, $M_{\rm tot}$ vs $e$, and $M_{\rm tot}$ vs $i$ -- those parameters have small degeneracies. Nevertheless, we have a strong constraint on the total mass $M_{\rm tot}=$\bestmtot. We rule out alternative orbit models with larger total mass $M_{\rm tot} > 65$~$M_{J}$ by starting multiple MCMC runs, as detailed in~\S\ref{sec:mcmcall}, with initial guesses of $M_{\rm tot}=70-120~M_{J}$. We find in all of these runs that the zero temperature walkers for $M_{\rm tot}$ steadily move towards lower masses ($\sim62~M_{J}$) over the duration of the run.  

As shown in Figure~\ref{fig:triangleabs}, proper motions $\mu_{\alpha}$, $\mu_{\delta}$ and reference epoch R.A. $\alpha_{0}$ and Decl $\delta_{0}$ are degenerate with mass ratio $q$. Nevertheless, we constrain the mass ratio to ~$q=$\massratio, in good agreement with the mass ratio of $q=0.78\pm0.10$ from \cite{Sahlmann15}, which uses only the VLT/FORS2 2013-2014 astrometry included in this analysis. Our relative parallax of $\varpi_{\rm rel}=$\para~arcseconds is $\sim1.5\sigma$ off from the relative parallax of $\varpi_{\rm rel}=0.50023\pm0.00011$ arcseconds measured by~\cite{Sahlmann15}. This discrepancy likely results from the addition of new 2015 FORS2 data, 2014-2015 GeMS data, and explicit astrometric analysis of the 1984 ESO plate in our analysis. Our median proper motions from the posterior probability density functions of $\mu_{\alpha}=$\pmra~arcseconds per year and $\mu_{\delta}=$\pmdec~arcseconds per year are in good agreement with \cite{Sahlmann15} proper motions of $\mu_{\alpha}=2.754\pm0.00625$~arcseconds per year and $\mu_{\delta}=0.3587\pm0.0095$~arcseconds per year. 

For earlier runs of the MCMC analysis, we placed much wider constraints of $\pm200$~mas on CDAR parameters $\rho_{\rm FORS2}$ and $d_{\rm FORS2}$, finding the two parameters to be tightly correlated (see Figure 3 from \citet{Lazorenko07}, which uses different units).  This linear relationship between $\rho_{\rm FORS2}$ and $d_{\rm FORS2}$ is shown in our Figure~\ref{fig:triangleabs}. This trend is expected as the LADC parameter $d_{\rm FORS2}$ models a near-cancellation of atmospheric DAR, as modeled by $\rho_{\rm FORS2}$. We narrowed the priors for these parameters to physically meaningful values as calculated from equation (4).  These limits are given in Table~\ref{table:paramrange} and reflected in Figure~\ref{fig:triangleabs}. \cite{Sahlmann15} find CDAR parameters $\rho_{\rm FORS2}=35.23$ mas and $d_{\rm FORS2}=-48.71$ mas which agree with the linear relationship between those parameters shown in Figure~\ref{fig:triangleabs}.  Furthermore, we find $\rho_{\rm GeMS}$ to be nearly zero, which agrees with our expectation that the usage of a narrowband $J_{\rm cont}$ filter minimizes the effect \citep{Cameron09}. 

In this study, we explored the possibility that the orbit model and the barycentric motion model could have degenerate parameters, given that both are derived from the same data sets. From Figure~\ref{fig:trianglecross}, however, we find no degeneracies between the orbit parameters and the barycenter astrometry parameters -- all of the 2D posterior PDFs are circularly symmetric.  Figure~\ref{fig:trianglemass} shows that the posterior PDFs for $q$ and $M_{\rm tot}$ are circularly symmetric, indicating little degeneracy between the mass of the primary $q$ and M$_{\rm tot}$. We do find a small degeneracy between the mass of the primary $M_A$ and secondary $M_B$, which is incorporated into our confidence intervals. These masses are within range of the model-based masses of $21-62$~$M_{J}$ estimated from the observed luminosities and effective temperatures of Luhman 16 AB and an upper limit of $<62$~$M_{J}$ (for both stars) from lithium detection by \cite{Faherty14} and \cite{Lodieu15}. 

In panels B and C of Figure~\ref{fig:minchi2}, we compare $10^4$ randomly sampled orbits from our zero temperature walker posterior probability distributions to our CDAR-subtracted separation data. The standard deviation of the absolute value of the FORS2 and GeMS residuals (``data-model'') is~$\approx$\bestsigmarel~mas for the separation fits as shown in panels D, E and F of Figure~\ref{fig:minchi2}. From panel D, we see a small systematic offset of $\approx0.5$~mas between the mean residuals for FORS2 orbital astrometry (red points) and the GeMS astrometry (teal points). This small offset may be due to small systematic errors in computation of the FORS2 CDAR parameters - however, the overall uncertainty on total mass (\bestmtoterr)~remains small. In multiple MCMC runs with and without the 1984 ESO and 1999 DENIS astrometry, we find the combination of both of these epochs plays a crucial role in constraining a mass-eccentricity-period degeneracy observed when using only the GeMS and FORS2 data to fit the orbital parameters of Luhman 16 AB. When using only the GeMS and FORS2 astrometry, we find that dynamical mass could only be constrained to 60-90 $M_{J}$. Observations in 2017-2018 correspond to time of closest separation for Luhman 16, which could render the DENIS 1999 data unnecessary. Similarly, observations in 2020-2022 would constrain the most uncertain parts of the orbit, possibly rendering the ESO 1984 data unnecessary. 

In Figure~\ref{fig:relrv}, we compare the same $10^{4}$ randomly sampled orbits to the CRIRES relative radial velocity points, finding good agreement. The relative radial velocities had little effect on further constraining the total mass beyond the constraints provided by the orbital astrometry. Using the RV data, we rule out orbits with argument of periastron $\omega > 150^{\circ}$, longitude of ascending node $\Omega < 129.5$. Those ranges of parameter space could not be ruled out by the astrometry alone.

In Figure~\ref{fig:absresid}, we compare randomly sampled barycenter models from our posterior probability distributions to the observed barycenter motion for Luhman 16 AB. We find residuals of~$1.41$~mas in R.A. and $0.99$ in declination for our barycentric astrometry with no statistically significant systematic trends. 
 
\subsection{Comparison to \cite{Bedin17} Results}

\cite{Bedin17} present Hubble Space Telescope Wide Field Camera 3 (WFC3) measurements of the separation of the individual components of Luhman 16 AB.  The authors also present derivations of the orbit and barycentric motion parameters of the system.  Due to the limited temporal coverage of the data relative to the full orbital period, the constraints on the orbit, total system mass, and individual masses are relatively weak.  Here we compare our measured separation data for the 2014.5 - 2015.5 epoch, in which both of our observational programs have coverage.  We also compare our derived model parameters.  

Our best-fitting model of the Luhman 16 AB orbit compares favorably to the raw separation measurements from \cite{Bedin17}.  The standard deviations of the residuals (i.e., our minimum $\chi^{2}$ model minus separation data from \cite{Bedin17} over the longest common epoch of 2014.5 - 2015.5) are $0.318$ milliarcseconds in X and $0.431$ milliarcseconds in Y, which compare favorably with the $0.3-0.4$ measurement errors given in Table 8 of that work.

All six barycentric motion parameters (parallax $\varpi_{\rm rel}$, proper motion $\mu_{\alpha,\delta}$, position $\alpha_0$, $\delta_0$, and mass ratio $q$) measured by \cite{Bedin17} fall within our $68.3\%$ confidence intervals.  The widths of our $68.3\%$ confidence intervals are generally similar to the $1\sigma$ error bars quoted by \cite{Bedin17} for these parameters.  

For the orbital parameters, given our more complete sampling of the orbital period, the widths of our confidence intervals are generally improved over the \citet{Bedin17} $1\sigma$ error bars.  Our measurements of the total mass $M_{\rm tot} = M_A+M_B$, semi-major axis $a$, eccentricity $e$, longitude of ascending node~$\Omega$, argument of periastron $\omega$, time of periastron passage $\tau$, inclination $i$, and period $P$ are more precise by factors of $2-30$.  Although our measurements of the longitude of ascending node~$\Omega$ and argument of periastron $\omega$ are different from those of \citet{Bedin17} by a $\pi$ phase shift and the inclination $i$ has an additional negative sign, our measurements of all orbital parameters are consistent to within the $1\sigma$ error bars quoted in \citet{Bedin17}, with the exception of the eccentricity $e$.  Our measurement of this parameter has a $68.3\%$ confidence interval of $[0.33, 0.37]$ and the value from \citet{Bedin17} has confidence limits of $[0.399, 0.527]$, so the difference is $1.76\sigma$ discrepant from our most likely value of $0.35$.  Given that this is the only parameter with a discrepancy above $1\sigma$, and the probability of such a difference occurring by chance alone is not small, we conclude that all of our orbital parameters are consistent to within the errors.

\subsection{Evolutionary Models}

Our masses and the bolometric luminosities of Luhman 16 AB due to \citep{Lodieu15} and \citep{Faherty14} are shown in Table~\ref{table:masslum}, and plotted in Figure~\ref{fig:evol} (we adopt the Lodieu et al. luminosities).
We compare our observations to cloudy (f$_{\rm sed}=2$), cloud-free, and hybrid (f$_{\rm sed}=2$ clouds-to-cloud-free) evolutionary models from \cite{Saumon08} (SM08). 

The evolutionary models use solar metallicity ($\frac{M}{H}=0$). The uncertainty on the bolometric luminosities include the known few percent photometric variability of both brown dwarfs \citep[see][for details]{Faherty14,Lodieu15}. We find the hybrid models give a common age of $\approx600-800$~Myr for both Luhman 16 AB to within $1\sigma$ uncertainty. Similar to the findings of \cite{Dupuy15} for L-T transition binary SDSSJ 1052 AB, we find a mass-luminosity relationship across the L-T transition consistent with zero. We also find that SM08 cloudless models cannot reproduce our masses and the observed luminosities of Luhman 16 AB to within $1\sigma$. Stronger conclusions to differentiate between the SM08 hybrid and SM08 cloudy models require greater precision in the bolometric luminosities for Luhman 16 AB.

\section{Conclusion} 

In this work, we derive the individual masses for the closest known brown dwarf binary, Luhman 16 AB. We draw upon archival observations from the European Southern Observatory Schmidt Telescope, the Deep Near-Infrared Survey of the Southern Sky (DENIS), public FORS2 data on the Very Large Telescope (VLT), and new astrometry from the Gemini South Multiconjugate Adaptive Optics System (GeMS).  We use an MCMC analysis to simultaneously fit the barycentric space motion of the binary as well as the mutual Keplerian orbit.  We see little evidence for degeneracies between parameters derived from the separation data and those describing the barycentric motion.  Our individual masses derived for Luhman 16 AB of~\masspri~$M_{J}$ and~\masssec~$M_{J}$~respectively represent the second derivation of masses for the L and T components of a dwarf binary with uncertainties $<10\%$. 

Luhman 16 AB mass measurements will also play an important role in constraining mass-luminosity relationship across the L/T transition. Future work may find our individual mass measurements of Luhman 16 AB to complement the observations by \cite{Dupuy15} of L-T transition binary SDSS J105213.51$+$442255.7, which found that hybrid cloudy-to-cloud-free brown dwarf evolutionary models of \citep{Saumon08} pass the co-evality test of binarity. On the other hand, cloudless models of \citet{Saumon08} are unable to reproduce the observed masses and luminosities of SDSS J1052+44 while maintaining co-evality, i.e. derive ages for both stars from the masses and luminosities that are within 1$\sigma$ uncertainty. Future work includes comparing the masses and luminosities of Luhman 16 AB to every set of the most recent cloudy and cloud-free evolutionary models, to test whether the models can reproduce the observed masses and luminosities. 


\begin{table}[htbp]
\footnotesize
\begin{center}
Table~\ref{table:absastro}: Position Measurements for Luhman 16 AB \\
\begin{tabular}{cccccccc}
\hline
Instrument & MJD        & $\alpha_{1}$     & $\delta_{1}$ & $\alpha_{2}$ & $\delta_{2}$ \\
           & (days)     & ($^{\circ}$, mas)     & ($^{\circ}$, mas)      & ($^{\circ}$, mas)       & ($^{\circ}$, mas) \\
ESO Schmidt (Red)& $45764.95$ & $162.3487979\pm179.51$ & $-53.3212708\pm169.87$ & $162.3482743\pm233.46$ & $-53.3208797\pm226.13$ \\
FORS2/VLT& $56397.00$ & $162.3111711\pm1.65$ & $-53.3182805\pm1.42$ & $162.3106864\pm1.65$ & $-53.3180021\pm1.46$ \\
	& $56402.06$ & $162.3111447\pm1.52$ & $-53.3182687\pm1.36$ & $162.3106612\pm1.53$ & $-53.3179911\pm1.36$ \\
	& $56408.04$ & $162.3111090\pm1.53$ & $-53.3182555\pm1.34$ & $162.3106269\pm1.52$ & $-53.3179791\pm1.34$ \\
	& $56418.03$ & $162.3110542\pm1.72$ & $-53.3182314\pm1.53$ & $162.3105750\pm1.70$ & $-53.3179576\pm1.65$ \\
	& $56424.09$ & $162.3110300\pm2.33$ & $-53.3182171\pm1.99$ & $162.3105528\pm2.25$ & $-53.3179446\pm1.93$ \\
	& $56432.97$ & $162.3109810\pm2.24$ & $-53.3181949\pm1.45$ & $162.3105053\pm1.69$ & $-53.3179242\pm1.48$ \\
	& $56438.03$ & $162.3109655\pm1.53$ & $-53.3181826\pm1.36$ & $162.3104909\pm1.51$ & $-53.3179129\pm1.37$ \\
	& $56443.99$ & $162.3109410\pm1.89$ & $-53.3181675\pm1.45$ & $162.3104679\pm1.81$ & $-53.3178995\pm1.45$ \\
	& $56448.00$ & $162.3109277\pm1.54$ & $-53.3181581\pm1.35$ & $162.3104557\pm1.56$ & $-53.3178908\pm1.34$ \\
	& $56452.97$ & $162.3109107\pm1.50$ & $-53.3181461\pm1.32$ & $162.3104397\pm1.50$ & $-53.3178798\pm1.32$ \\
	& $56457.00$ & $162.3109010\pm1.58$ & $-53.3181373\pm1.44$ & $162.3104316\pm1.58$ & $-53.3178717\pm1.44$ \\
	& $56459.97$ & $162.3108918\pm1.59$ & $-53.3181301\pm1.43$ & $162.3104225\pm1.55$ & $-53.3178656\pm1.38$ \\
	& $56465.96$ & $162.3108774\pm1.49$ & $-53.3181169\pm1.78$ & $162.3104097\pm1.55$ & $-53.3178537\pm2.28$ \\
	& $56693.25$ & $162.3103384\pm1.51$ & $-53.3182177\pm1.35$ & $162.3099316\pm1.53$ & $-53.3180095\pm1.37$ \\
	& $56702.19$ & $162.3102721\pm1.52$ & $-53.3182173\pm1.32$ & $162.3098669\pm1.51$ & $-53.3180112\pm1.32$ \\
	& $56726.14$ & $162.3100969\pm1.55$ & $-53.3182010\pm1.35$ & $162.3097001\pm1.56$ & $-53.3180015\pm1.38$ \\
	& $56735.19$ & $162.3100353\pm1.52$ & $-53.3181895\pm1.37$ & $162.3096414\pm1.62$ & $-53.3179926\pm1.34$ \\
	& $56746.12$ & $162.3099561\pm1.56$ & $-53.3181720\pm1.35$ & $162.3095644\pm1.64$ & $-53.3179776\pm1.33$ \\
	& $56757.11$ & $162.3098827\pm1.52$ & $-53.3181511\pm1.33$ & $162.3094951\pm1.53$ & $-53.3179596\pm1.35$ \\
	& $56773.00$ & $162.3097794\pm1.51$ & $-53.3181165\pm1.34$ & $162.3093965\pm1.51$ & $-53.3179293\pm1.35$ \\
	& $56782.99$ & $162.3097251\pm1.82$ & $-53.3180926\pm1.31$ & $162.3093454\pm1.88$ & $-53.3179084\pm1.78$ \\
	& $56795.98$ & $162.3096628\pm1.54$ & $-53.3180604\pm1.31$ & $162.3092875\pm1.60$ & $-53.3178793\pm1.32$ \\
	& $57013.31$ & $162.3092730\pm1.68$ & $-53.3180357\pm1.48$ & $162.3089733\pm2.39$ & $-53.3179157\pm1.70$ \\
	& $57018.33$ & $162.3092495\pm1.63$ & $-53.3180422\pm1.35$ & $162.3089518\pm1.70$ & $-53.3179239\pm1.38$ \\
	& $57025.34$ & $162.3092125\pm1.50$ & $-53.3180504\pm1.34$ & $162.3089167\pm1.51$ & $-53.3179339\pm1.34$ \\
	& $57034.35$ & $162.3091610\pm1.53$ & $-53.3180597\pm1.33$ & $162.3088688\pm1.56$ & $-53.3179459\pm1.35$ \\
	& $57044.36$ & $162.3090986\pm1.58$ & $-53.3180676\pm1.42$ & $162.3088113\pm1.67$ & $-53.3179570\pm1.53$ \\
GeMS& $56701.22$ & $162.3102827\pm1.72$ & $-53.3182209\pm1.72$ & $162.3098775\pm1.72$ & $-53.3180149\pm1.72$ \\
	& $56759.18$ & $162.3098694\pm1.72$ & $-53.3181510\pm1.72$ & $162.3094820\pm1.72$ & $-53.3179602\pm1.72$ \\
	& $56760.17$ & $162.3098617\pm1.72$ & $-53.3181480\pm1.72$ & $162.3094745\pm1.72$ & $-53.3179576\pm1.72$ \\
	& $56804.07$ & $162.3096275\pm1.72$ & $-53.3180436\pm1.72$ & $162.3092542\pm1.72$ & $-53.3178649\pm1.72$ \\
	& $57000.34$ & $162.3093359\pm1.72$ & $-53.3180168\pm1.72$ & $162.3090309\pm1.72$ & $-53.3178932\pm1.72$ \\
	& $57086.32$ & $162.3087965\pm1.72$ & $-53.3180638\pm1.72$ & $162.3085243\pm1.72$ & $-53.3179656\pm1.72$ \\
\hline
\end{tabular}
\caption{Observed astrometry for Luhman 16 AB measured as described in \S\ref{sect:relastro}. The CDAR is not subtracted from these data.}
\label{table:absastro}
\end{center}
\end{table}

\begin{table}[htbp]
\footnotesize
\begin{center}
Table~\ref{table:relastro}: Orbital Astrometry \\
\begin{tabular}{cccc}
\hline
Instrument & MJD        & $\Delta$X      & $\Delta$Y        \\
           & (days)     & (mas)  & (mas)\\
ESO Schmidt (Red)& $45764.95$ & $-1125.97\pm160.00$ & $1408.10\pm135.00$ \\
FORS2/VLT& $56397.00$ & $-1042.56\pm1.41$ & $1002.32\pm2.03$ \\
	& $56402.06$ & $-1039.95\pm0.47$ & $999.61\pm0.81$ \\
	& $56408.04$ & $-1036.84\pm0.47$ & $995.03\pm1.02$ \\
	& $56418.03$ & $-1030.59\pm1.10$ & $986.03\pm2.30$ \\
	& $56424.09$ & $-1026.21\pm0.65$ & $981.09\pm1.63$ \\
	& $56432.97$ & $-1023.17\pm0.81$ & $974.80\pm1.60$ \\
	& $56438.03$ & $-1020.58\pm0.66$ & $970.98\pm1.12$ \\
	& $56443.99$ & $-1017.40\pm0.86$ & $965.15\pm1.37$ \\
	& $56448.00$ & $-1014.90\pm0.65$ & $962.26\pm1.07$ \\
	& $56452.97$ & $-1012.86\pm0.53$ & $958.55\pm0.99$ \\
	& $56457.00$ & $-1009.39\pm0.64$ & $956.42\pm1.39$ \\
	& $56459.97$ & $-1009.20\pm1.06$ & $952.25\pm0.69$ \\
	& $56465.96$ & $-1005.63\pm0.78$ & $947.62\pm1.98$ \\
	& $56693.25$ & $-874.99\pm1.44$ & $749.51\pm2.28$ \\
	& $56702.19$ & $-871.52\pm0.89$ & $741.83\pm1.68$ \\
	& $56726.14$ & $-853.54\pm2.36$ & $718.19\pm3.34$ \\
	& $56735.19$ & $-847.04\pm1.08$ & $709.08\pm1.35$ \\
	& $56746.12$ & $-842.25\pm1.00$ & $699.73\pm1.29$ \\
	& $56757.11$ & $-833.66\pm1.24$ & $689.32\pm1.95$ \\
	& $56773.00$ & $-823.64\pm0.81$ & $673.97\pm1.25$ \\
	& $56782.99$ & $-816.79\pm0.97$ & $663.35\pm2.68$ \\
	& $56795.98$ & $-807.20\pm1.24$ & $652.02\pm1.32$ \\
	& $57013.31$ & $-644.66\pm1.85$ & $432.15\pm1.25$ \\
	& $57018.33$ & $-640.39\pm2.46$ & $426.14\pm1.95$ \\
	& $57025.34$ & $-636.18\pm1.33$ & $419.43\pm1.68$ \\
	& $57034.35$ & $-628.30\pm1.11$ & $409.74\pm1.78$ \\
	& $57044.36$ & $-617.60\pm1.34$ & $398.22\pm2.12$ \\
GeMS& $56701.22$ & $-871.33\pm0.26$ & $741.46\pm0.28$ \\
	& $56759.18$ & $-833.18\pm0.26$ & $686.77\pm0.28$ \\
	& $56760.17$ & $-832.71\pm0.26$ & $685.66\pm0.28$ \\
	& $56804.07$ & $-802.82\pm0.26$ & $643.57\pm0.28$ \\
	& $57000.34$ & $-655.87\pm0.26$ & $444.92\pm0.28$ \\
	& $57086.32$ & $-585.38\pm0.26$ & $353.41\pm0.28$ \\
DENIS& $51220.50$ & $<675.00$ & $<675.00$ \\
\hline
\end{tabular}
\caption{Observed separations for Luhman 16 AB measured as described in \S\ref{sect:relastro}. We present an upper limit on the separation at the epoch of the DENIS observations as Luhman 16 AB is unresolved. The maximum-likelihood CDAR is subtracted from these orbital astrometry observations, as detailed in \S\ref{sec:cdarmod} and \S\ref{sec:orb}.}
\label{table:relastro}
\end{center}
\end{table}

\begin{table}[htbp]
\begin{center}
Table~\ref{table:relrv}: CRIRES Relative Radial Velocity \\
\begin{tabular}{cc}
\hline
MJD        & $\Delta$V      \\
(days)     & (m s$^{-1}$)  \\
56417.5 & 2740$\pm$200 \\
56779.5 & 1940$\pm$200 \\
56797.5 & 1850$\pm$200 \\
\hline
\end{tabular}
\caption{VLT/CRIRES observed relative radial velocities of Luhman 16 AB used in fitting the orbital parameters as detailed in \S\ref{sec:crires}.}
\label{table:relrv}
\end{center}
\end{table}

\begin{table}[htbp]
\begin{center}
Table~\ref{table:paramrange}: Uniform priors for our MCMC analysis 
\begin{tabular}{lll}
\hline 
\hline
Parameter & Min & Max \\
\hline
\hline
\multicolumn{3}{c}{Orbital Parameters}\\
\hline
Semi-major Axis, $a$ (AU) & $2.0$ & $50.0$ \\
Total Mass, $M_{\rm tot}$ ($M_{J}$) & $10.5$ & $125.7$ \\
Eccentricity, $e$ & $0.0$ & $0.9$ \\
Inclination, $\cos{i}$ & $-0.9$ & $1.0$ \\
Longitude of Ascending Node, $\Omega$ ($^{\circ}$) & $0.0$ & $360.0$ \\
Argument of Periastron, $\omega$ ($^{\circ}$) & $0.0$ & $360.0$ \\
Time of Periastron Passage, $\tau$ (day) & $10000$ & $80000$ \\
\hline
\multicolumn{3}{c}{Barycenter Parameters}\\
\hline
Mass Ratio, $q$ & $0.2$ & $1.5$ \\
Ref R.A., $\alpha_0$ ($^{\circ}$) & $162.310622$ & $162.311178$ \\
Ref Decl., $\delta_0$ ($^{\circ}$) & $-53.318413$ & $-53.317857$ \\
Proper Motion R.A., $\mu_{\alpha}$ ($^{\prime\prime}$) & $-2.8$ & $-2.7$ \\
Proper Motion Decl., $\mu_{\delta}$ ($^{\prime\prime}$) & $0.3$ & $0.4$ \\
Parallax, $\varpi_{\rm rel}$ ($^{\prime\prime}$) & $0.49$ & $0.51$ \\
\hline
\multicolumn{3}{c}{Instrument Parameters}\\
\hline
CDAR FORS2, $\rho_{\rm FORS2}$ (mas) & $25.0$ & $50.0$ \\
CDAR FORS2, $d_{\rm FORS2}$ (mas) & $-58.9$ & $-38.5$ \\
CDAR GeMS, $\rho_{\rm GeMS}$ (mas) & $0.0$ & $5.0$ \\
\end{tabular}
\end{center}


\caption{Uniformly sampled prior probability density distributions 
used in our MCMC analysis (see~\S\ref{sec:mcmcall}).
$^{1}$ CDAR is Chromatic Differential Atmospheric Refraction. }

\label{table:paramrange}
\end{table}

\begin{table}[htbp]
\begin{center}
Table~\ref{table:bestfit}: MCMC Posterior Probability Distributions \\
\begin{tabular}{lllll}
\hline 
\hline
Parameter & Sample Fit$^2$ & Median & 68.3\% c.i. & 95.4\% c.i. \\
\hline
\multicolumn{5}{c}{Orbital Parameters}\\
\hline
Orbital Period, $P$ (yr)& $27.4$ & $27.4$ & $27.6$,$27.1$ & $27.6$,$26.7$ \\
Semimajor Axis, $a$ (AU) & $3.55$ & $3.54$ & $3.59$,$3.49$ & $3.64$,$3.43$ \\
Total Mass, $M_{\rm tot}$ ($M_{J}$)& $62.5$ & $62.0$ & $64.1$,$60.3$ & $66.2$,$58.9$ \\
Eccentricity, $e$ & $0.35$ & $0.35$ & $0.37$,$0.33$ & $0.39$,$0.31$ \\
Inclination, $i$ ($^{\circ}$)& $79.6$ & $79.5$ & $79.3$,$79.8$ & $79.0$,$80.1$ \\
Longitude of Ascending Node, $\Omega$ ($^{\circ}$) & $130.17$ & $130.12$ & $130.24$,$130.00$ & $130.34$,$129.87$ \\
Argument of Periastron, $\omega$ ($^{\circ}$)& $129.6$ & $130.4$ & $133.9$,$126.9$ & $137.6$,$123.6$ \\
Time of Periastron Passage, $\tau$ (day)& $48050$ & $48030$ & $48340$,$47780$ & $48720$,$47580$ \\
\hline
\multicolumn{5}{c}{Barycenter Parameters}\\
\hline
Mass Ratio, $q$ & $0.82$ & $0.82$ & $0.85$,$0.78$ & $0.88$,$0.75$ \\
Ref R.A., $\alpha_0$ ($^{\circ}$) & $162.31093$ & $162.31093$ & $162.31093$,$162.31092$ & $162.31094$,$162.31092$ \\
Ref Decl., $\delta_0$ ($^{\circ}$) & $-53.31815$ & $-53.31815$ & $-53.31814$,$-53.31815$ & $-53.31814$,$-53.31815$ \\
Proper Motion R.A., $\mu_{\alpha}$ ($^{\prime\prime}$) & $-2.761$ & $-2.762$ & $-2.760$,$-2.765$ & $-2.757$,$-2.767$ \\
Proper Motion Decl., $\mu_{\delta}$ ($^{\prime\prime}$) & $0.358$ & $0.358$ & $0.361$,$0.354$ & $0.365$,$0.351$ \\
Parallax, $\varpi_{\rm rel}$ ($^{\prime\prime}$) & $0.50150$ & $0.50101$ & $0.50153$,$0.50048$ & $0.50204$,$0.49995$ \\
\hline
\multicolumn{5}{c}{Instrument Parameters}\\
\hline
CDAR$^1$ FORS2, $\rho_{\rm FORS2}$ (mas) & $29.92$ & $38.06$ & $46.09$,$29.74$ & $49.34$,$26.13$ \\
CDAR FORS2, $d_{\rm FORS2}$ (mas) & $-42.70$ & $-48.73$ & $-41.93$,$-55.32$ & $-39.06$,$-58.04$ \\
CDAR GeMS, $\rho_{\rm GeMS}$ (mas) & $0.14$ & $0.49$ & $1.12$,$0.13$ & $1.94$,$0.02$ \\
\end{tabular}


\caption{Luhman 16 AB parameters in our MCMC analysis (see~\S\ref{sec:mcmcall}).
$^{1}$ CDAR is Chromatic Differential Atmospheric Refraction. 
$^{2}$ Sample fit is taken directly from the accepted zero temperature walker MCMC chain. The sample 16-parameter fit is chosen to be within the 68.3\% confidence interval.  
The median parameters are not representative of a single actual 16 parameter fit (i.e. set of parameters accepted by the MCMC sampler). 
}

\label{table:bestfit}
\end{center}
\end{table}

\begin{table}[htbp]
\footnotesize
\begin{center}
Table~\ref{table:cdar}: Chromatic Differential Atmospheric Refraction Parameters \\
\begin{tabular}{ccccccc}
\hline
Instrument & MJD        & f$_{3m}$     & $\tan{z_{\rm m}}$ & $\tan{z_{\rm L,m}}$ & $\cos{\gamma}$ & $\sin{\gamma}$ \\
FORS2/VLT& $56396.9951520$ & $0.9791000$ & $0.7546000$ & $0.9286000$ & $-0.6232000$ & $-0.7816000$ \\
	& $56402.0627130$ & $0.9915000$ & $0.5493000$ & $0.6744000$ & $-0.9973000$ & $-0.0604000$ \\
	& $56408.0405880$ & $0.9809000$ & $0.5521000$ & $0.6636000$ & $-0.9910000$ & $-0.1282000$ \\
	& $56418.0314980$ & $0.9882000$ & $0.5505000$ & $0.6652000$ & $-0.9941000$ & $0.0876000$ \\
	& $56424.0852670$ & $0.9971000$ & $0.7135000$ & $0.8958000$ & $-0.6789000$ & $0.7336000$ \\
	& $56432.9703430$ & $1.0054000$ & $0.5540000$ & $0.6837000$ & $-0.9855000$ & $-0.1509000$ \\
	& $56438.0316910$ & $1.0052000$ & $0.6561000$ & $0.8121000$ & $-0.7741000$ & $0.6323000$ \\
	& $56443.9929650$ & $1.0101000$ & $0.5929000$ & $0.7314000$ & $-0.8962000$ & $0.4426000$ \\
	& $56447.9978380$ & $0.9892000$ & $0.6350000$ & $0.7693000$ & $-0.8124000$ & $0.5820000$ \\
	& $56452.9719010$ & $0.9842000$ & $0.6028000$ & $0.7154000$ & $-0.8769000$ & $0.4730000$ \\
	& $56456.9999390$ & $1.0011000$ & $0.7344000$ & $0.9070000$ & $-0.6485000$ & $0.7599000$ \\
	& $56459.9735320$ & $0.9914000$ & $0.6628000$ & $0.8087000$ & $-0.7625000$ & $0.6462000$ \\
	& $56465.9621110$ & $1.0012000$ & $0.6804000$ & $0.8611000$ & $-0.7322000$ & $0.6810000$ \\
	& $56693.2499970$ & $0.9981000$ & $0.5626000$ & $0.6850000$ & $-0.9664000$ & $-0.2378000$ \\
	& $56702.1862400$ & $0.9887000$ & $0.6508000$ & $0.7781000$ & $-0.7880000$ & $-0.6114000$ \\
	& $56726.1444170$ & $0.9856000$ & $0.5884000$ & $0.7062000$ & $-0.9101000$ & $-0.4007000$ \\
	& $56735.1855440$ & $0.9945000$ & $0.5759000$ & $0.6892000$ & $-0.9348000$ & $0.3368000$ \\
	& $56746.1216870$ & $0.9966000$ & $0.5497000$ & $0.6716000$ & $-0.9968000$ & $-0.0488000$ \\
	& $56757.1135790$ & $0.9882000$ & $0.5595000$ & $0.6750000$ & $-0.9730000$ & $0.2073000$ \\
	& $56772.9970680$ & $0.9914000$ & $0.6361000$ & $0.7689000$ & $-0.8148000$ & $-0.5743000$ \\
	& $56782.9852030$ & $0.9883000$ & $0.5937000$ & $0.7066000$ & $-0.8974000$ & $-0.4399000$ \\
	& $56795.9827440$ & $0.9915000$ & $0.5522000$ & $0.6678000$ & $-0.9914000$ & $-0.0771000$ \\
	& $57013.3143040$ & $0.9976000$ & $0.7249000$ & $0.8769000$ & $-0.6650000$ & $-0.7466000$ \\
	& $57018.3266580$ & $0.9896000$ & $0.6318000$ & $0.7469000$ & $-0.8224000$ & $-0.5647000$ \\
	& $57025.3373360$ & $0.9899000$ & $0.5669000$ & $0.6854000$ & $-0.9567000$ & $-0.2760000$ \\
	& $57034.3479320$ & $0.9853000$ & $0.5530000$ & $0.6668000$ & $-0.9876000$ & $0.1321000$ \\
    & $57044.3577360$ & $0.9873000$ & $0.6153000$ & $0.7421000$ & $-0.8512000$ & $0.5177000$ \\
GeMs	& $56701.2201260$ & $0.9642000$ & $0.4498000$ & $0.0000000$ & $-0.9218000$ & $-0.3867000$ \\
	& $56759.1760450$ & $0.9890000$ & $0.6395000$ & $0.0000000$ & $-0.5429000$ & $0.8396000$ \\
	& $56760.1718400$ & $0.9747000$ & $0.6243000$ & $0.0000000$ & $-0.5675000$ & $0.8224000$ \\
	& $56804.0736350$ & $0.9871000$ & $0.7281000$ & $0.0000000$ & $-0.4182000$ & $0.9083000$ \\
	& $57000.3432270$ & $0.9695000$ & $0.6273000$ & $0.0000000$ & $-0.5403000$ & $-0.8413000$ \\
	& $57086.3244570$ & $0.9653000$ & $0.8635000$ & $0.0000000$ & $-0.2703000$ & $0.9625000$ \\
\hline
\end{tabular}
\caption{Chromatic differential atmospheric refraction parameters used to calculate parameters $f_{\rm 1,x,m}$, $f_{\rm 1,y,m}$, $f_{\rm 2,x,m}$, and $f_{\rm 2,y,m}$ as detailed in \S\ref{sec:cdarmod}. $\tan{z_{L,m}}$ is zero for the GeMS data given that GeMS does not have a atmospheric differential corrector like VLT/FORS2. }
\label{table:cdar}
\end{center}
\end{table}

\begin{table}[htbp]
\begin{center}
Table~\ref{table:masslum}: Masses and Luminosities of Luhman 16 AB \\
\begin{tabular}{cccc}
\hline
\hline
 & Mass ($M_{J}$) & $\log{L_{\odot}}^1$ & $\log{L_{\odot}}^2$ \\
\hline
Luhman 16 A & $34.2\pm1.2$ & $-4.67\pm0.04$ & $-4.66\pm0.08$ \\
Luhman 16 B & $27.9\pm1.0$ & $-4.71\pm0.10$ & $-4.68\pm0.13$ \\
\end{tabular}
\caption{$^1$ \cite{Faherty14}.
$^2$ \cite{Lodieu15}.
}
\label{table:masslum}
\end{center}
\end{table}

\begin{figure}[ht!]
\begin{center}
	\includegraphics[width = \textwidth]{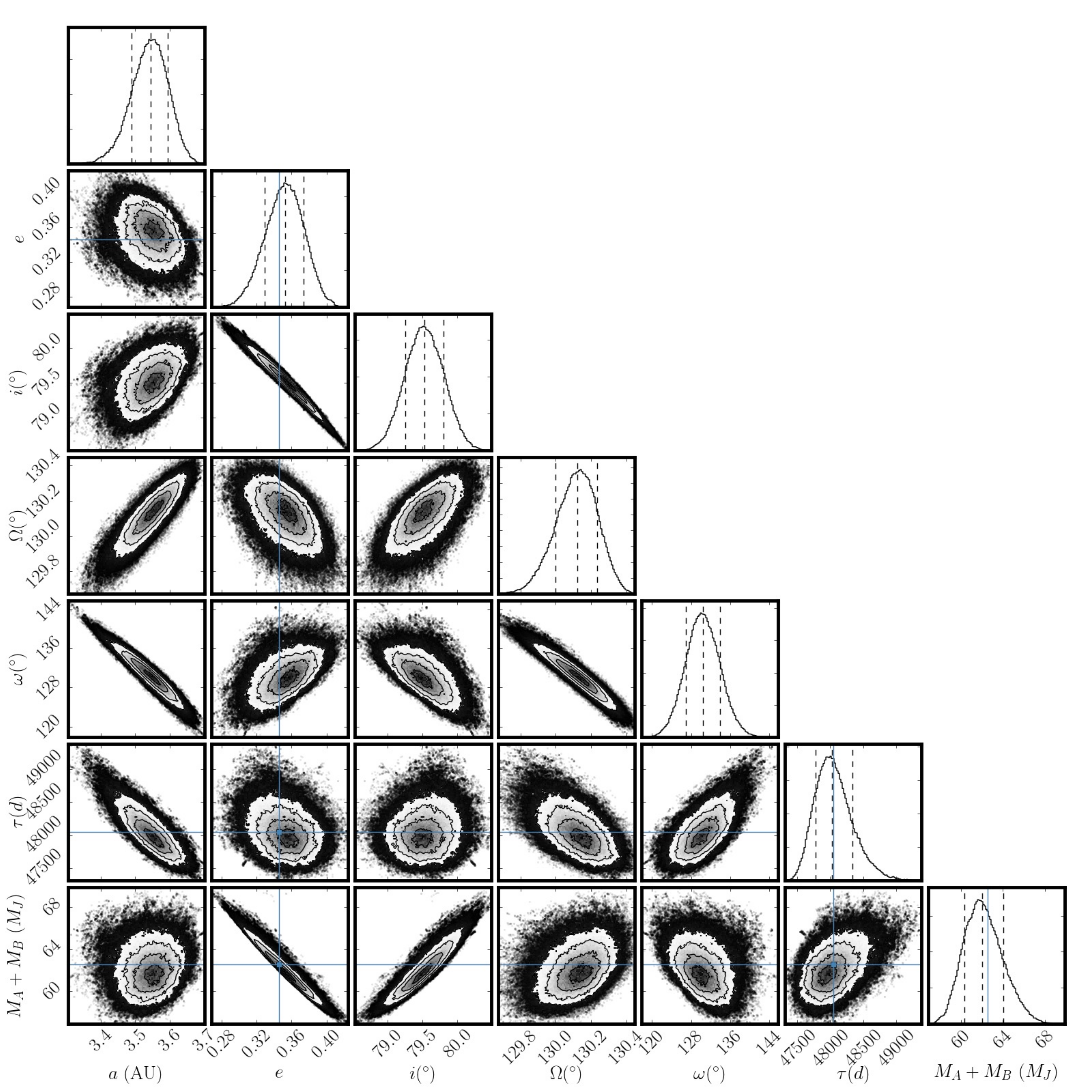}

	\caption{Posterior probability distributions for the orbital astrometry parameters (see \S\ref{sec:mcmcall} and Table~\ref{table:bestfit}). There are clear degeneracies between several of the parameters ($\omega$,$e$,$a$,$\Omega$ and $M_{\rm tot}$). However, the confidence intervals on the total mass $M_{\rm tot}$ given the degeneracies remain small at~$\approx$\bestmtoterr. 
	}
	\label{fig:trianglerel}
\end{center}
\end{figure}

\begin{figure}[ht!]
\begin{center}
	\includegraphics[width = \textwidth]{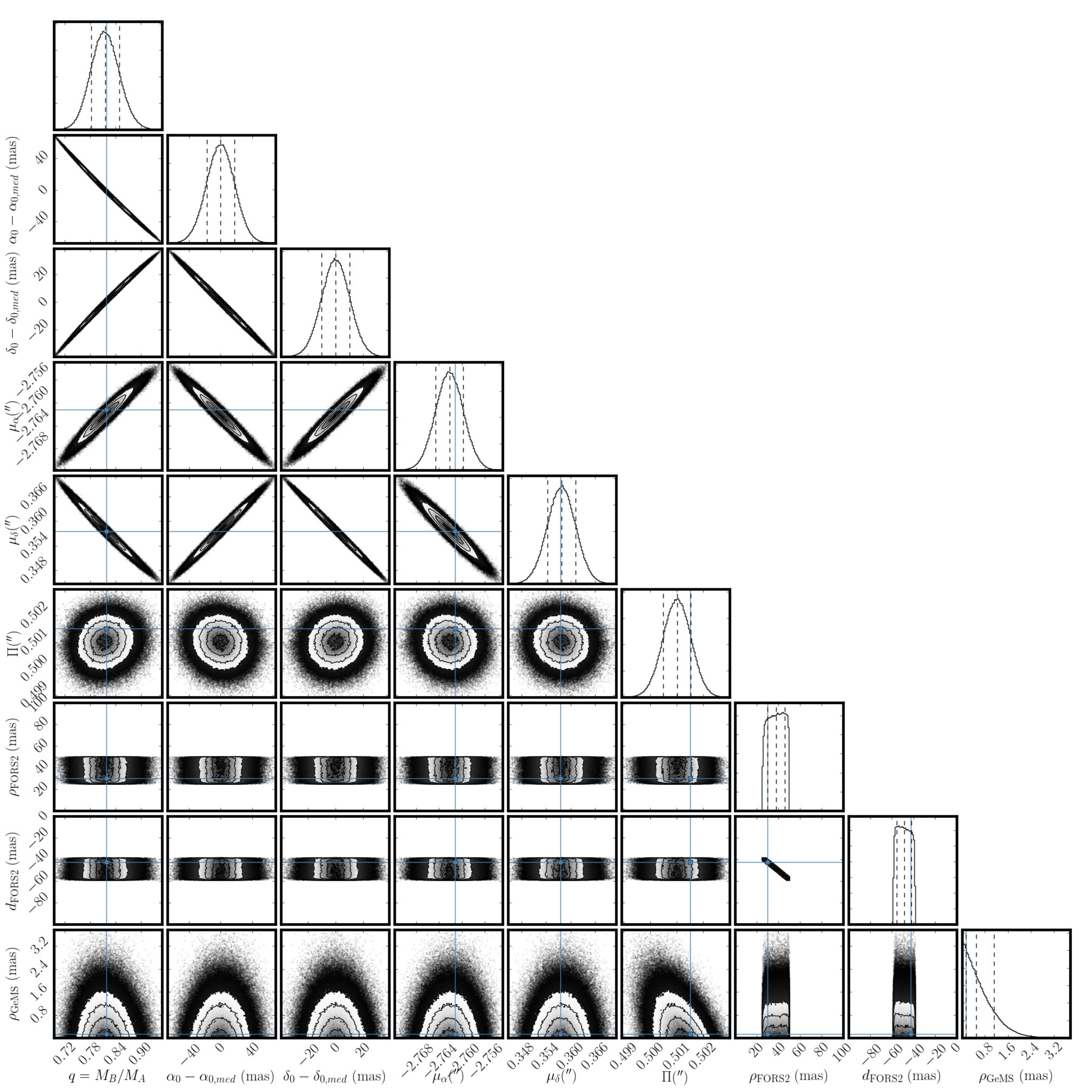}

	\caption{Posterior probability distributions for the barycentric astrometry and instrument parameters (see \S\ref{sec:mcmcall} and Table~\ref{table:bestfit}). There are obvious degeneracies between several of the parameters, in particular between proper motion $\mu_{\alpha}$,$\mu_{\delta}$ and the mass ratio $q$. Nevertheless, our mass ratio~\massratio~is in good agreement with $0.78\pm0.10$ from \cite{Sahlmann15}. 
	}
	\label{fig:triangleabs}
\end{center}
\end{figure}

\begin{figure}[ht!]
\begin{center}
    \includegraphics[trim={0 0 48.9cm 38.2cm},clip,width = \textwidth]{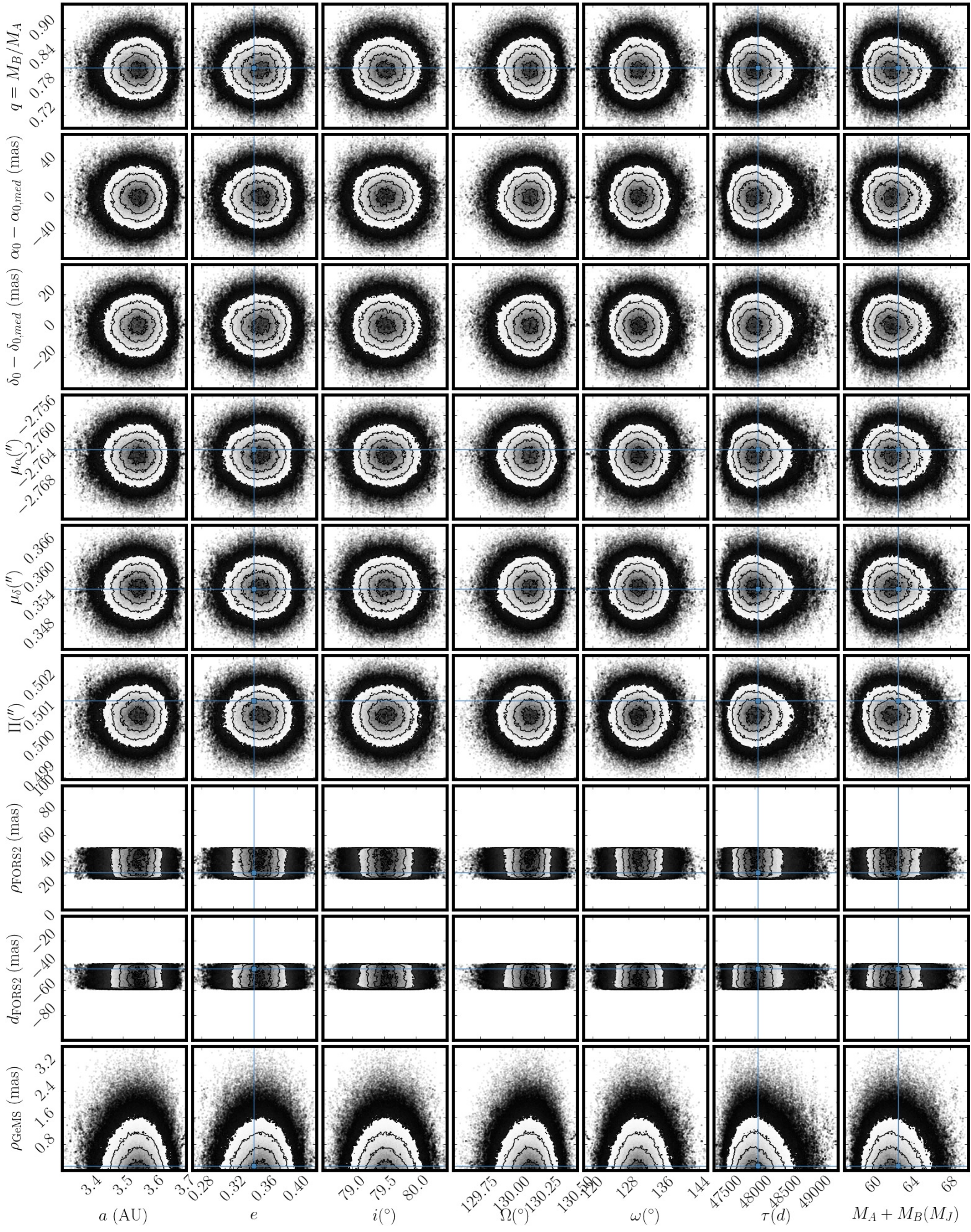}

	\caption{Posterior probability distributions from our MCMC analysis for the cross section between our orbital astrometry parameters (X-axis) and or barycentric astrometry parameters (Y-axis). There are no obvious degeneracies between the parameters derived from Luhman 16 separation data and the barycentric motion parameters.
	}
	\label{fig:trianglecross}
\end{center}
\end{figure}

\begin{figure}[ht!]
\begin{center}
	\includegraphics[width = \textwidth]{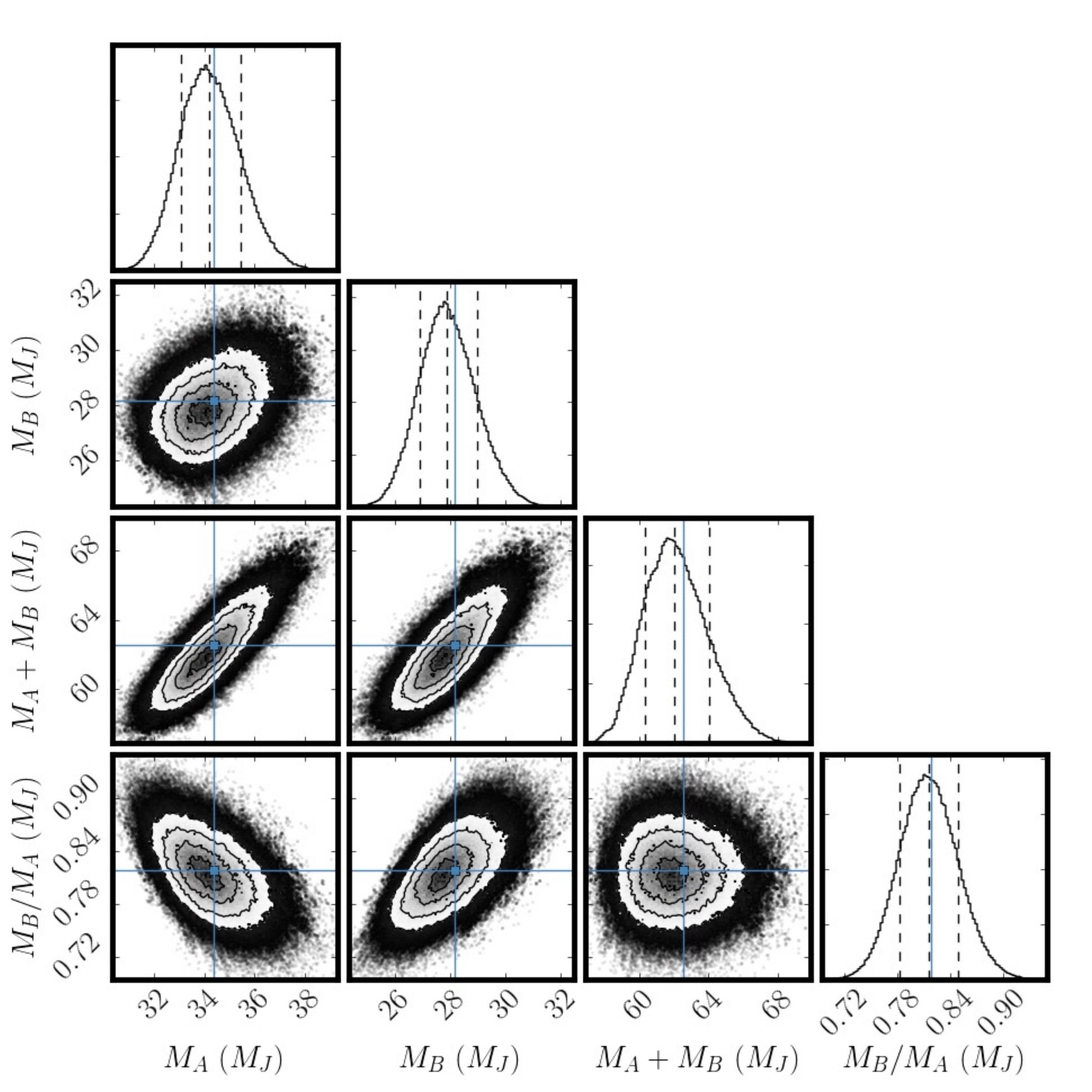}

	\caption{Posterior probability distributions for the mass ratio, total mass, and individual masses (see \S\ref{sec:mcmcall} and Table~\ref{table:bestfit}). The individual masses $M_A$ and $M_B$ are largely decorrelated. We marginalize over the posterior probability distributions to obtain a well-constrained total mass $M_{\rm tot}=$\bestmtot~$M_{\odot}$ with a~\bestmtoterr (1$\sigma$) uncertainty. 
	}
	\label{fig:trianglemass}
\end{center}
\end{figure}

\begin{figure}[ht!]
\begin{center}

	\includegraphics[width= \textwidth]{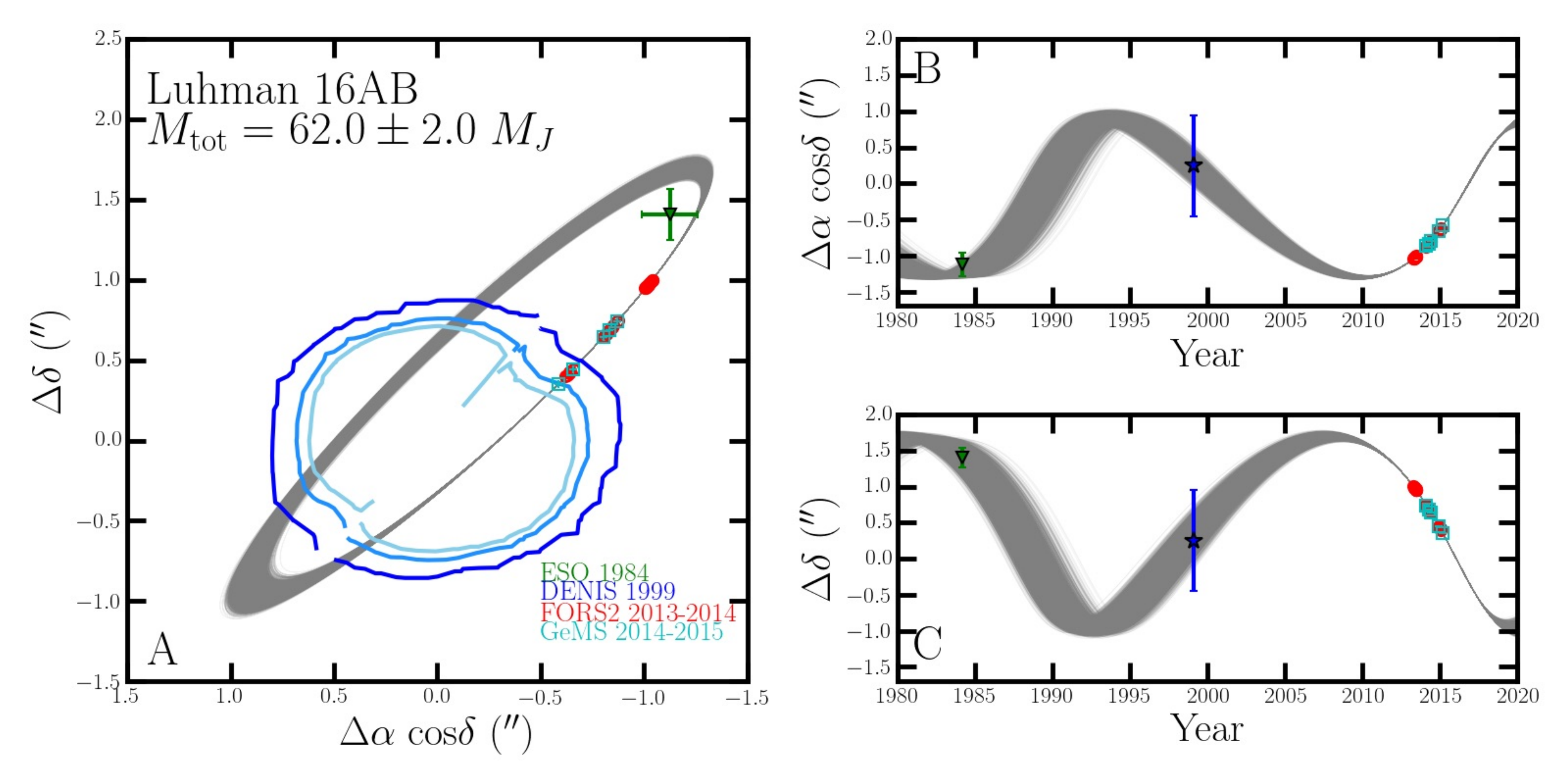}
	\includegraphics[width= \textwidth]{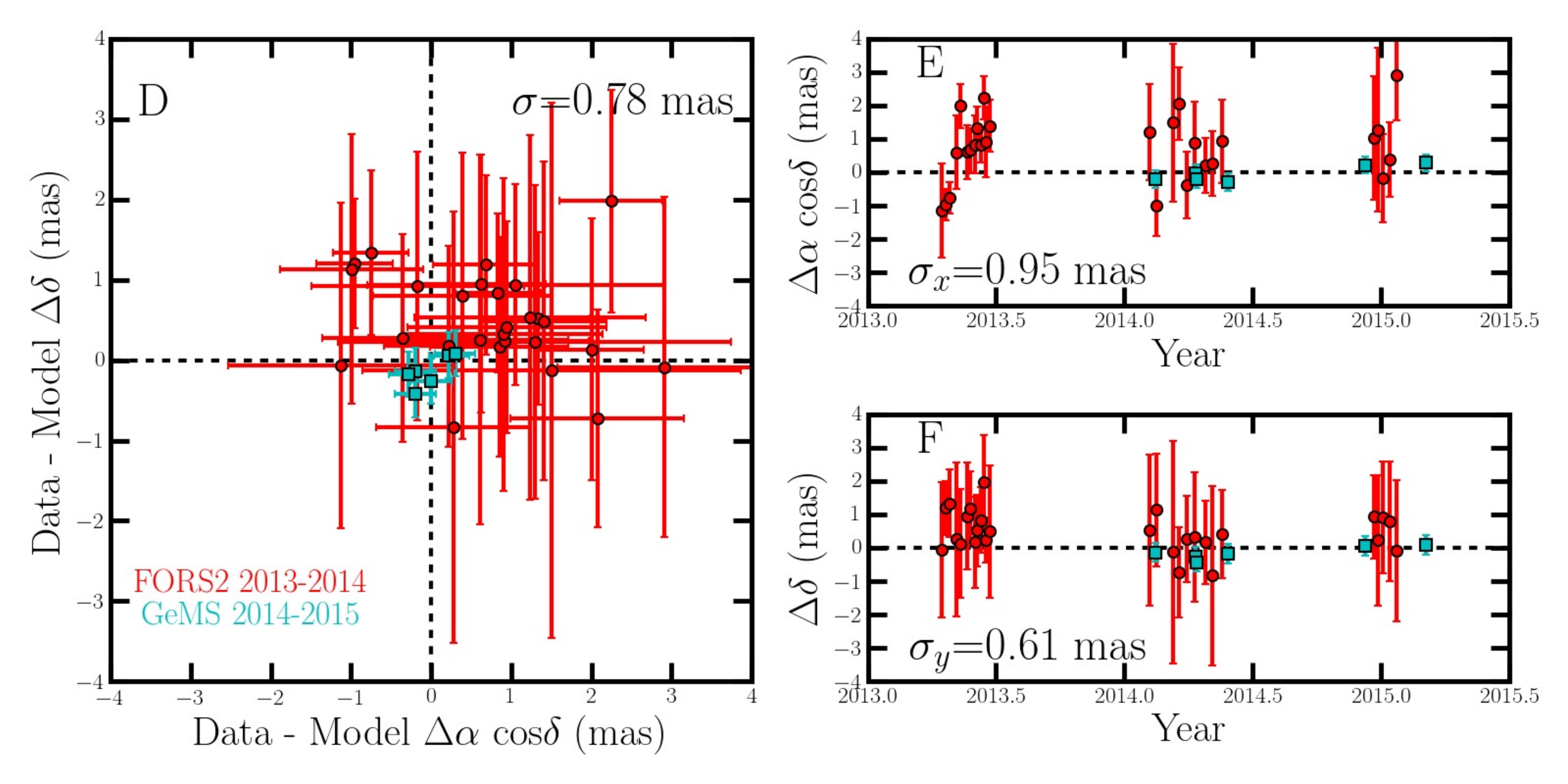}
	\caption{{\bf Panel A:} $10^{4}$ randomly sampled orbits from our posterior probability distributions (gray lines) using an MCMC analysis technique (see \S\ref{sec:orb}). The uncertainty for the FORS2 (red circles) and GeMS (teal squares) orbital astrometry are smaller than the symbol sizes. The $1\sigma$, $2\sigma$ and $3\sigma$ contours for our DENIS 1999 are shown in dark blue, blue, and light blue respectively. The dotted lines correspond to the 180 degree degeneracy in the binary fitting routine (see~\S\ref{sec:denis99}). The ESO 1984 epoch is shown as a green downward facing triangle. {\bf Panels B and C:} Mutual orbit of Luhman 16 AB in delta R.A. and Declination with the same models shown in Panel A. {\bf Panels D, E and F:} The ``Data-Model" residuals for our best fit orbit, excluding the 1984 and 1999 epochs for clarity. We find a small $\lesssim0.8$ mas systematic offset between our FORS2 and GeMS astrometry and our maximum-likelihood orbit, likely due to small systematic errors in the CDAR parameters (see~\S\ref{sec:cdarmod}). 
	}
	\label{fig:minchi2}
\end{center}
\end{figure}

\begin{figure}[ht!]
\begin{center}
	\includegraphics[width = \textwidth ]{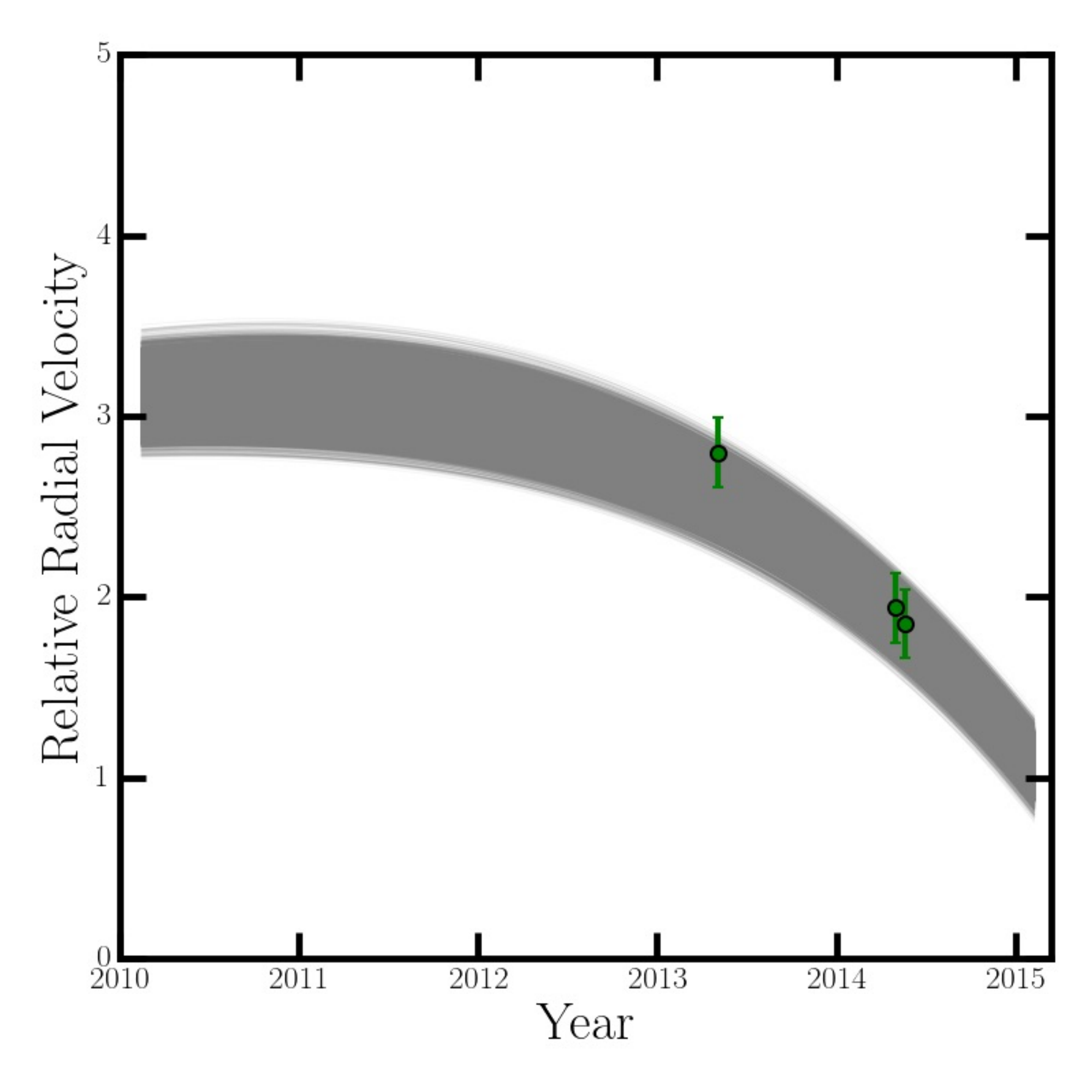}

	\caption{$10^{4}$ randomly sampled relative radial velocity orbits from our posterior probability MCMC chains (same as Figure~\ref{fig:minchi2}), compared to the VLT/CRIRES relative radial velocities $V_{A} - V_{B}$, as detailed in \S\ref{sec:crires} and \S\ref{sec:relrv}. The relative radial velocities are in good agreement with our maximum-likelihood orbit. 	
	}
	\label{fig:relrv}
\end{center}
\end{figure}

\begin{figure}[ht!]
\begin{center}
	\includegraphics[width = \textwidth ]{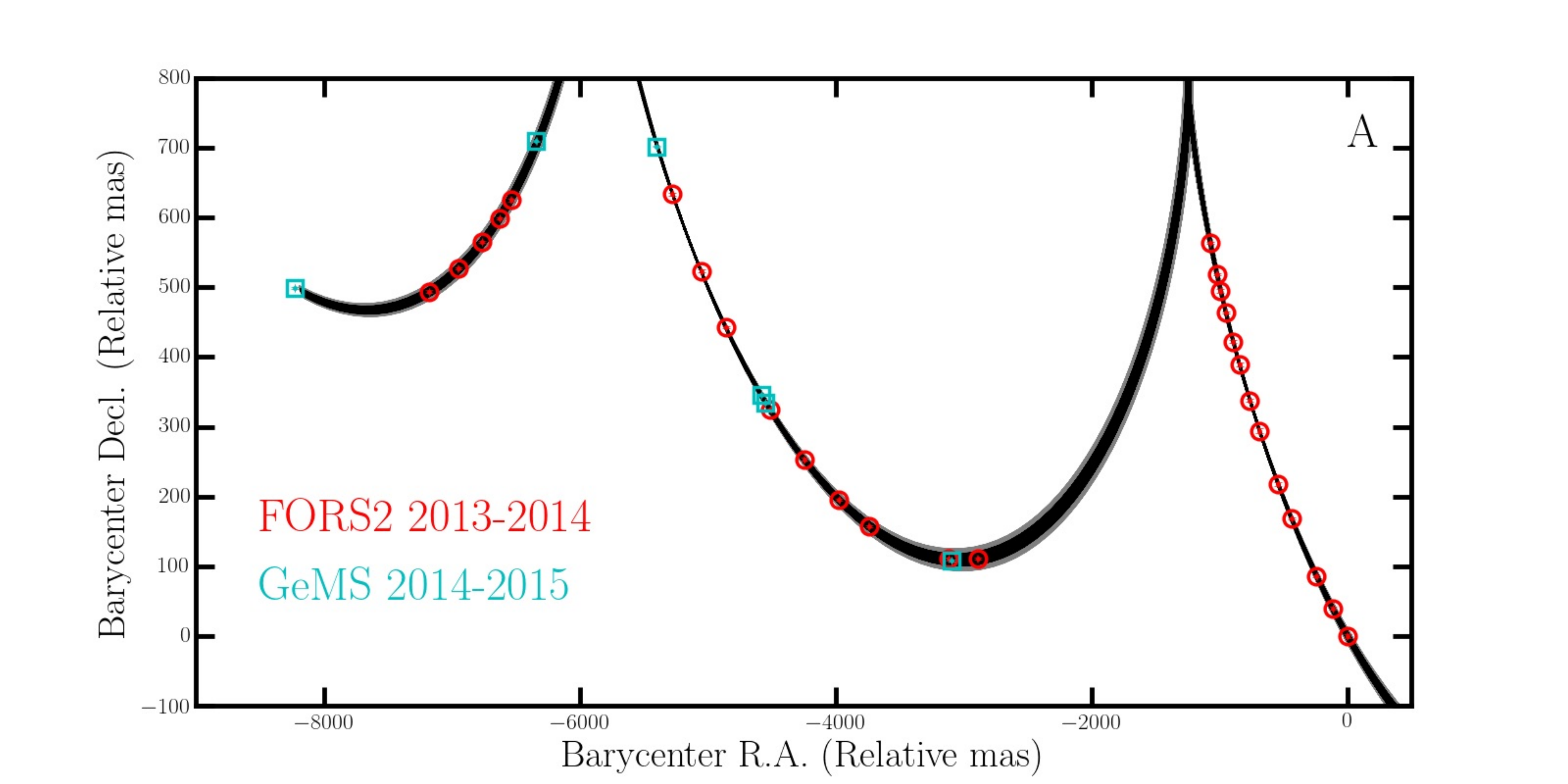}
	\includegraphics[ width = \textwidth ]{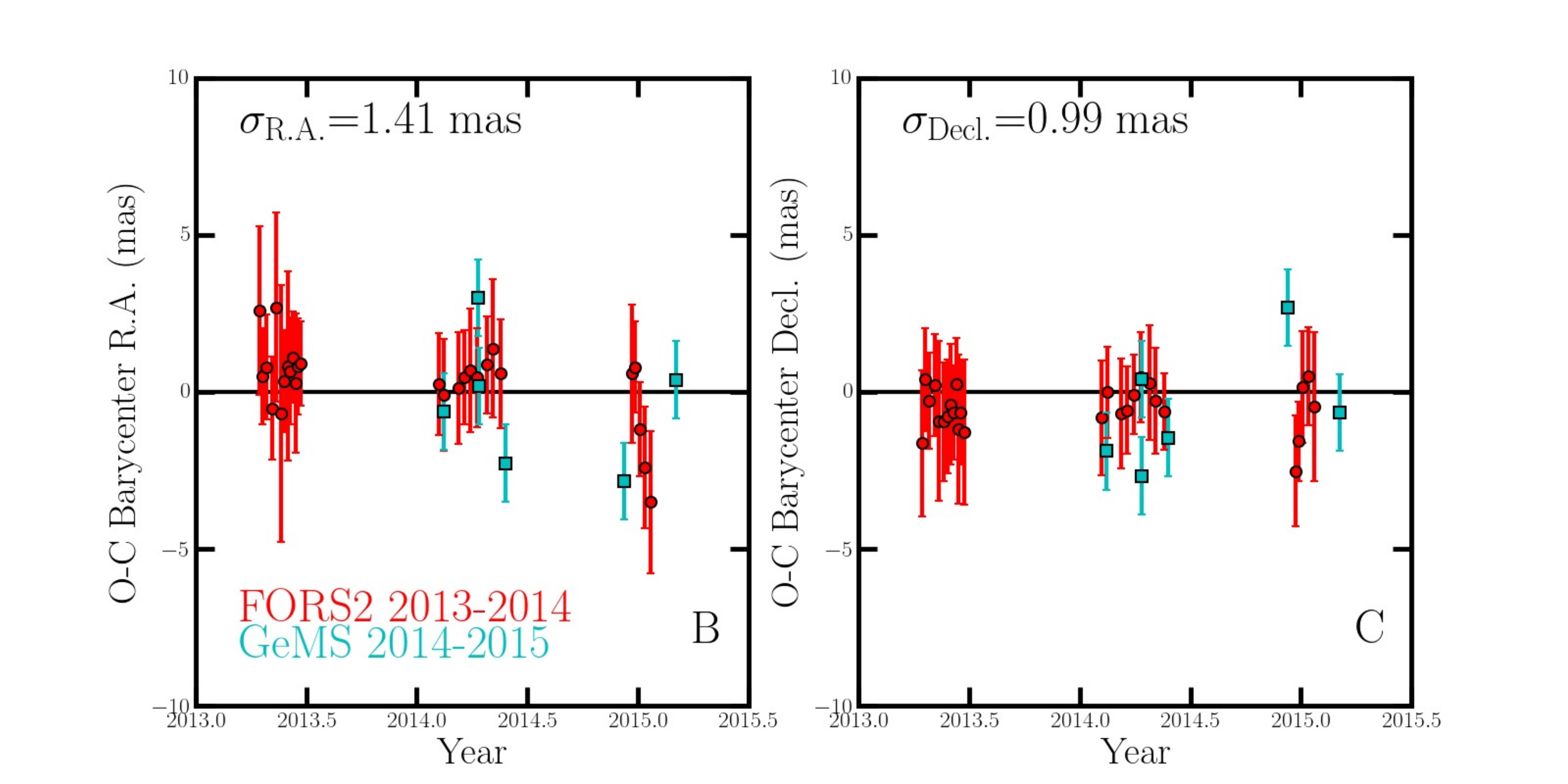}
	\caption{{\bf Panel A:} Our FORS2 and GeMS barycentric astrometry (red and teal circles and squares) are accurate enough to distinguish between different mass ratios. The black and grey lines are the barycenter locations corresponding MCMC chains within $1\sigma$ and $2\sigma$ confidence intervals of our maximum-likelihood mass ratio $q=$\massratio.  While the true location of the barycenter relative to either star does not change - the barycenter will vary as a function of the mass ratio (Equation~\ref{eqn:bary}, \S\ref{sec:abs}).  {\bf Panels B and C:} ``Data-Model" residuals for our maximum-likelihood barycentric astrometry, excluding the 1984 epoch for clarity. We find there is no systematic error between our GeMS (teal squares), FORS2 (red circles) astrometry and our astrometric model. 
	}
	\label{fig:absresid}
\end{center}
\end{figure}


\begin{figure}
\centering 
\subfloat{%
              \includegraphics[trim={3cm 2cm 3cm 3.8cm},clip,,width=0.55\textwidth]{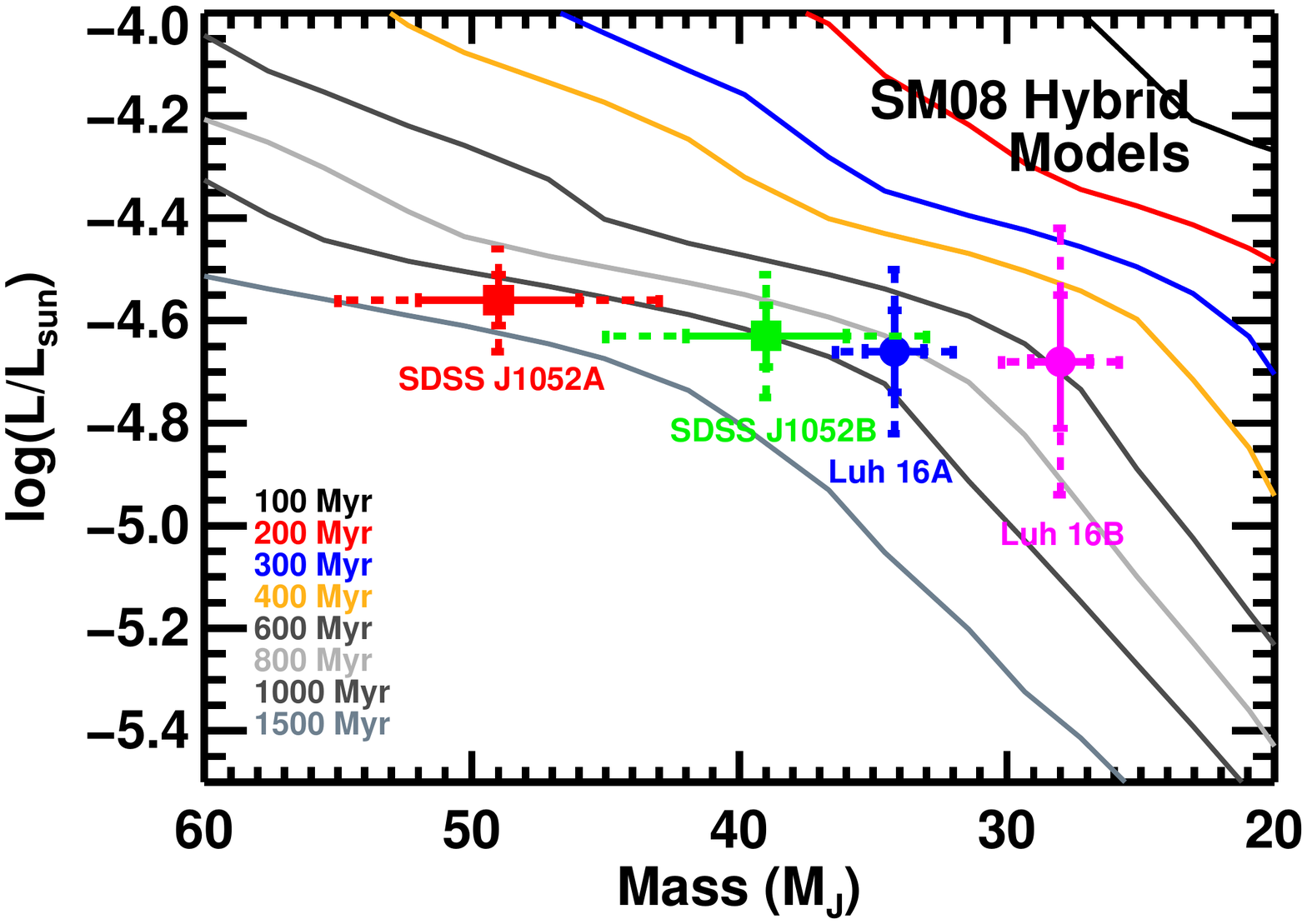}%
  \label{fig:sm08hybrid}%
}
\subfloat{%
              \includegraphics[trim={3cm 2cm 3cm 3.8cm},clip,,width=0.55\textwidth]{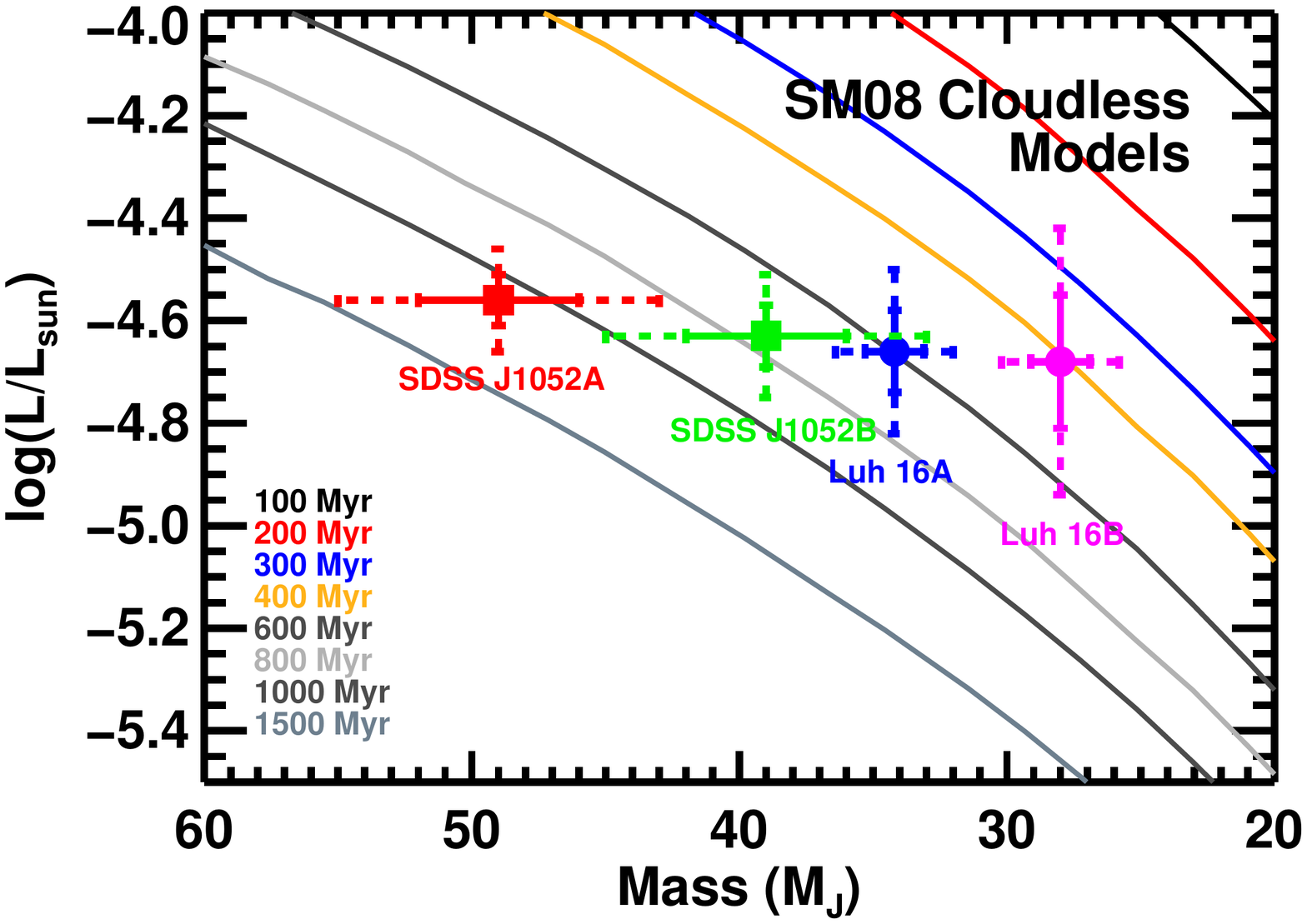}%
  \label{fig:sm08cloudless}%
}\qquad
\subfloat{%
            \includegraphics[trim={3cm 2cm 3cm 3.8cm},clip,,width=0.55\textwidth]{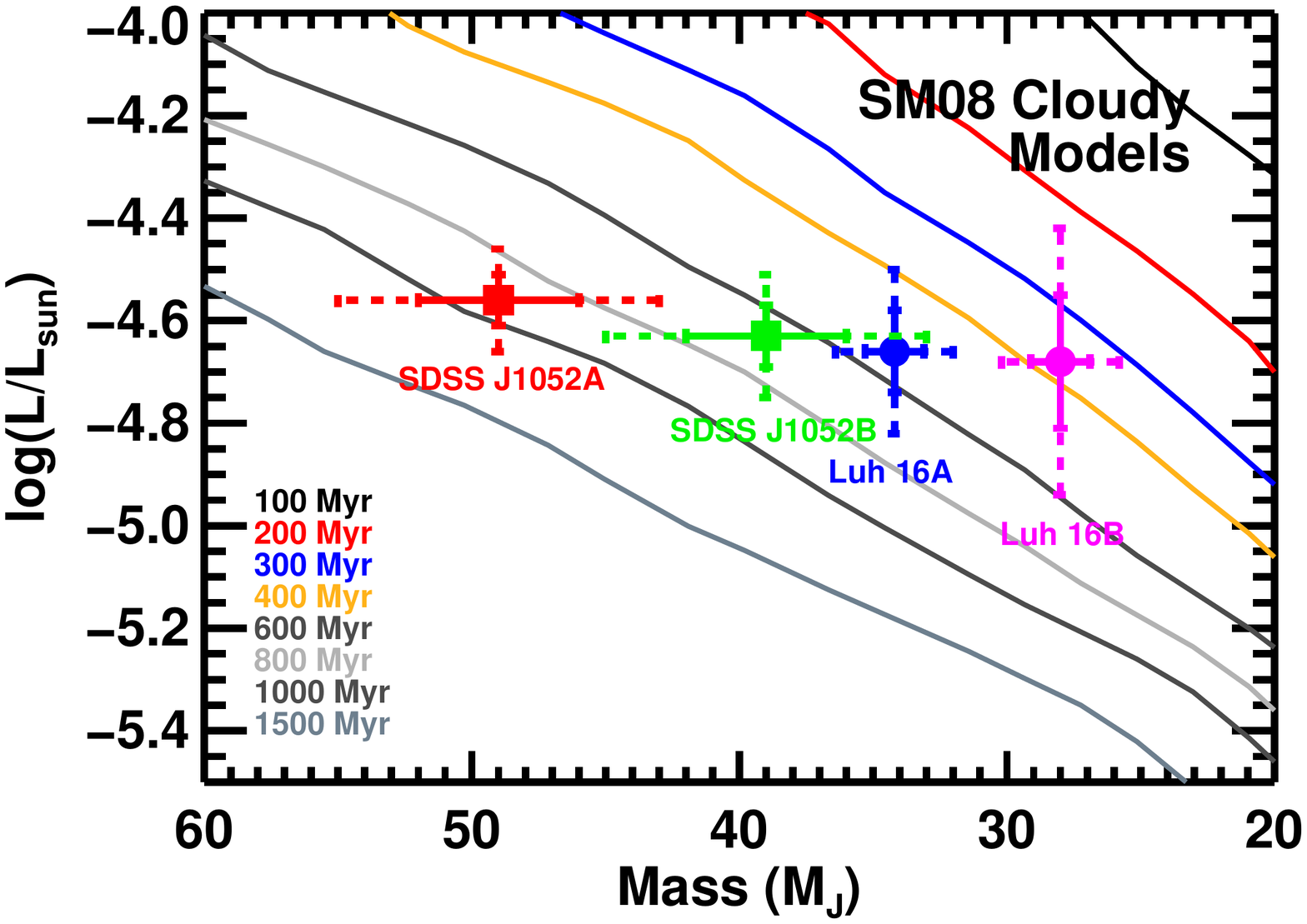}
            \label{fig:sm08cloudy} %
}
\subfloat{%
            \includegraphics[trim={3cm 2cm 3cm 3.8cm},clip,,width=0.55\textwidth]{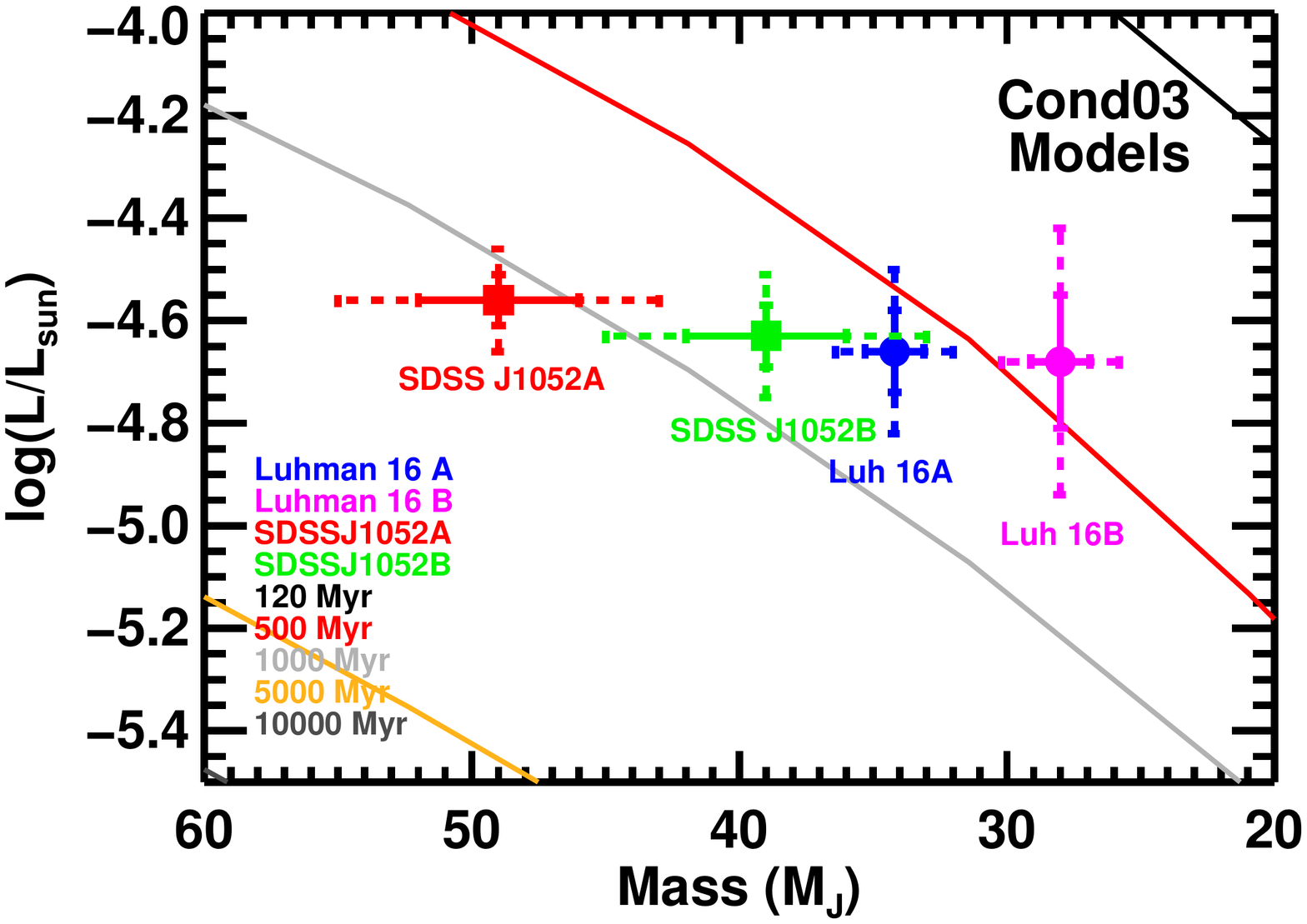}
            \label{fig:cond}%
}
        \caption[ CMD ]
        {\label{fig:evol} The hybrid (cloudy-to-cloud-free) or cloudy models of \cite{Saumon08}(SM08) best reproduce the flat mass-luminosity relationship across the L-T transition observed from directly measured masses and bolometric luminosities of brown dwarf binaries Luhman 16 AB \citep[for luminosities, this work for masses]{Lodieu15} and
        SDSSJ105213.51+442255.7AB \citep{Dupuy15} to within uncertainty. The uncertainties on the bolometric luminosity due to the few percent inherent photometric variability of Luhman 16 AB are included in the error bars shown here. 
        }
\end{figure}

\section{Acknowledgements}
Based on observations obtained at the Gemini Observatory,
which is operated by the Association of Universities
for Research in Astronomy, Inc., under a cooperative
agreement with the NSF on behalf of the Gemini
partnership: the National Science Foundation (United
States), the National Research Council (Canada), CONICYT
(Chile), the Australian Research Council (Australia),
Ministerio da Ciencia, Tecnologia e Inovacao
(Brazil) and Ministerio de Ciencia, Tecnologıa e Innovacion
Productiva (Argentina). Data are acquired through the
Gemini Science Archive and processed using the Gemini
IRAF package.  This work has made use of data from the European Space Agency (ESA)
mission GAIA~(\url{http://www.cosmos.esa.int/gaia}), processed by
the GAIA~Data Processing and Analysis Consortium (DPAC,
\url{http://www.cosmos.esa.int/web/gaia/dpac/consortium}). Funding
for the DPAC has been provided by national institutions, in particular
the institutions participating in the GAIA Multilateral Agreement.  This research is based on the data obtained from the ESO
Science Archive Facility under programme IDs 291.C-5004 and 593.C-0314.  This research has made use of data obtained from the SuperCOSMOS Science Archive, prepared and hosted by the Wide Field Astronomy Unit, Institute for Astronomy, University of Edinburgh, which is funded by the UK Science and Technology Facilities Council.  This work was partially funded by NASA/NEXSS NNX15AD95G.   This research has made use of the VizieR catalogue access tool, CDS, Strasbourg, France.  This work is performed under the auspices of the U.S. Department of Energy by Lawrence Livermore National Laboratory under Contract DE-AC52-07NA27344 with document release number LLNL-JRNL-701044. The posterior probability distributions in Figures~\ref{fig:trianglerel},\ref{fig:triangleabs},\ref{fig:trianglecross}, and \ref{fig:trianglemass} are made with \texttt{corner} \citep{Foreman16}.

\clearpage
\bibliography{bib}
\clearpage

\end{document}